\documentclass[aps,prd,twocolumn,nofootinbib]{revtex4}

\usepackage{amsmath}
\usepackage{graphicx}
\usepackage{dcolumn}
\usepackage{bm}
\usepackage{amssymb}
\usepackage{latexsym}
\usepackage{subfigure}

\usepackage{amsmath}
\usepackage{graphicx}
\usepackage{dcolumn}
\usepackage{bm}
\usepackage{amssymb}
\usepackage{latexsym}
\usepackage{subfigure}

\newcommand{\be}{\begin{equation}}
\newcommand{\ee}{\end{equation}}
\newcommand{\bq}{\begin{eqnarray}}
\newcommand{\eq}{\end{eqnarray}}

\begin{document}

\title{CMB Cold Spot from Inflationary Feature Scattering}
\date{\today}
\author{Yi Wang$^{1,2}$ and Yin-Zhe Ma$^{3,4}$}
\affiliation{
$^{1}$Department of Physics, The Hong Kong University of Science and Technology, Clear Water Bay, Kowloon, Hong Kong, PR China \\
$^{2}$Department of Applied Mathematics and Theoretical Physics, University of Cambridge, Cambridge CB3 0WA, UK \\
$^{3}$ School of Chemistry and Physics, University of KwaZulu-Natal, Westville Campus, Private Bag X54001, Durban, 4000, South Africa \\
$^{4}$Jodrell Bank Centre for Astrophysics, School of Physics and Astronomy, The University of Manchester, Oxford Road, Manchester M13 9PL, UK}

\begin{abstract}
We propose a ``feature-scattering'' mechanism to explain the cosmic microwave background cold spot seen from {\it WMAP} and {\it Planck} maps. If there are hidden features in the potential of multi-field inflation, the inflationary trajectory can be scattered by such features. The scattering is controlled by the amount of isocurvature fluctuations, and thus can be considered as a mechanism to convert isocurvature fluctuations into curvature fluctuations. This mechanism predicts localized cold spots (instead of hot ones) on the CMB. In addition, it may also bridge a connection between the cold spot and a dip on the CMB power spectrum at $\ell \sim 20$.
\end{abstract}

\maketitle

\section{Introduction}
\label{sec:intro}

\begin{figure}
  \centering
  \includegraphics[width=0.5\textwidth]{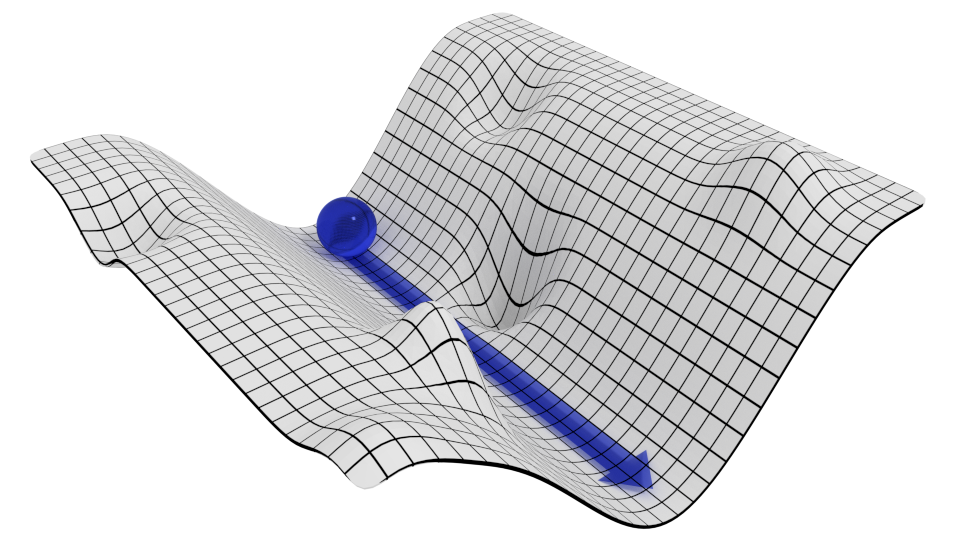}
  \caption{\label{fig:feature} An illustration of the feature scattering mechanism. The classical inflationary trajectory is featureless. However, because of the quantum fluctuation of the isocurvature direction, hidden features may be encountered. At such an encounter, the inflationary trajectory gets scattered, the inflaton losses kinetic energy and thus isocurvature fluctuation converts to curvature fluctuation. }
\end{figure}

In the recent years, observations of the cosmic microwave background (CMB) radiation fluctuations by the {\it Wilkinson Microwave Anisotropy Probe} ({\it WMAP}) and {\it Planck} satellite have led to a precise measurement of temperature fluctuations on the sky from the largest scales down to arcmin scales \cite{Hinshaw13,Planck16}. The temperature anisotropy is found to be highly Gaussian and ``statistically isotropic'' in the sense that nearly all statistical proprieties of the temperature anisotropy can be described by the angular power spectrum $C^{TT}_{\ell}$~\cite{Planck23}. However, it was found since {\it WMAP} 1-year data that there is a deep cold spot ($\Delta T \simeq -120\,$K) in the southern Galactic hemisphere along the direction ($l=209^{\rm o},\,b=-57^{\rm o}$) with angular radius $\theta \simeq 10^{\rm o}$~\cite{Vielva04,Cruz07}, which is further confirmed by {\it Planck} nominal mission data~\cite{Planck23}. The cold spot is highly non-Gaussian in the sense that the probability of the cold spot existing in the statistical isotropic Gaussian universe is less than $0.1$ percent~\cite{Gurzadyan14}.

Since then, the cold spot feature in the CMB map has invoked many observational and theoretical investigations. Initially, it was suggested that the unsubtracted foreground contamination might be responsible for the apparent non-Gaussian features~\cite{Chiang06,Tojeiro06}, but later studies~\cite{Cruz06,Planck23} showed that the significance of cold spot is not affected by Galactic residues in the region of the spot. It was also proposed that a spherically symmetric void with radius $\sim 300\,$Mpc at redshift $z=1$ can produce a large and deep CMB cold spot through the late-time integrated Sachs-Wolfe effect~(ISW)~\cite{Inoue06} (also known as the Rees-Sciama effect~\cite{Rees68}). Later studies~\cite{Szapudi14,Finelli14} with the galaxy survey data~\cite{Kovacs14} did find such a supervoid of size $r\simeq 195\,$Mpc with density contrast $\delta_{0}\simeq -0.1$ at redshift $z=0.16$ align with the cold spot direction. However, more detailed following-up studies~\cite{Nadathur14,Zibin14} showed that the Rees-Sciama effect produced by such a void is several orders of magnitude lower than the linear ISW effect therefore is not able to account for the observed feature.

The interesting non-Gaussian feature of cold-spot also invokes theorists to investigate the plausible explanation from the early universe. By considering various cosmological defects in the early universe, Refs.~\cite{Cruz07-sci,Cruz08} proposed that a cosmic ``texture'' (i.e. a concentration of stress-energy and a time-varying gravitational potential due to the symmetry-breaking phase transition) can generate hot and cold spots on the last-scattering surface, with the fundamental symmetry-breaking scale found to be $\phi_{0}\sim 10^{15}\,$GeV. However, by applying the Bayesian method to {\it WMAP} full-sky data, Ref.~\cite{Feeney12} did not find strong evidence of the texture model, neither completely rule out the possibility (at 95\% confidence level). It was also proposed that cosmic bubble collision, predicted by eternal inflation theories, can induce the density perturbation between our bubble and others, which can give arise to the localized features in the CMB~\cite{Aguirre07,Gurzadyan13}. But more detail data analysis~\cite{Feeney13} showed that the expected number of bubble collision is too few to account for the features in the CMB. Alternatively, a cold spot may follow from a different trajectory during multi-stream inflation~\cite{Li:2009sp, Afshordi:2010wn} or modulated reheating after multi-field inflation~\cite{Sanchez14}.

The above physical or astronomical interpretations of cold spot either fail at some level, or require fine tuning or exotic scenarios of the early universe. Economically, some cosmologists would prefer to interpret the cold spot merely as a ``$3\sigma$'' statistical fluke. In this paper, we will provide a natural and physically plausible explanation of the cold spot, through multiple-field inflation. If the inflationary trajectory is scattered by a feature hidden in the isocurvature direction, the inflaton loses some energy and thus inflation tends to be longer. Therefore, it is possible that only a small portion of the sky hits the feature due to stochastic fluctuations, then that local patch of the universe experiences longer period of inflation and thus produces a cold spot.  The mechanism is illustrated in Fig.~\ref{fig:feature}.

This paper is organized as follows. In Section~\ref{sec:cold-spot-from}, we provide an explicit example of feature scattering, from massless isocurvature directions. Our predictions are compared with the measurement of the cold-spot in {\it Planck}'s {\tt SMICA} map. In Section~\ref{sec:mass-isoc-direct}, we consider isocurvature directions with mass $m\sim H$. We conclude in Section~\ref{sec:concl-disc} and discuss possible future directions. Throughout the paper, the unit ${\rm M}_{\rm pl} = 1/\sqrt{8 \pi G} = 1$ is used unless otherwise stated.

\section{Cold spot from feature scattering}
\label{sec:cold-spot-from}
\subsection{Potential choice and method of calculation}

\begin{figure*}[htbp]
  \centering
  \includegraphics[width=0.53\textwidth]{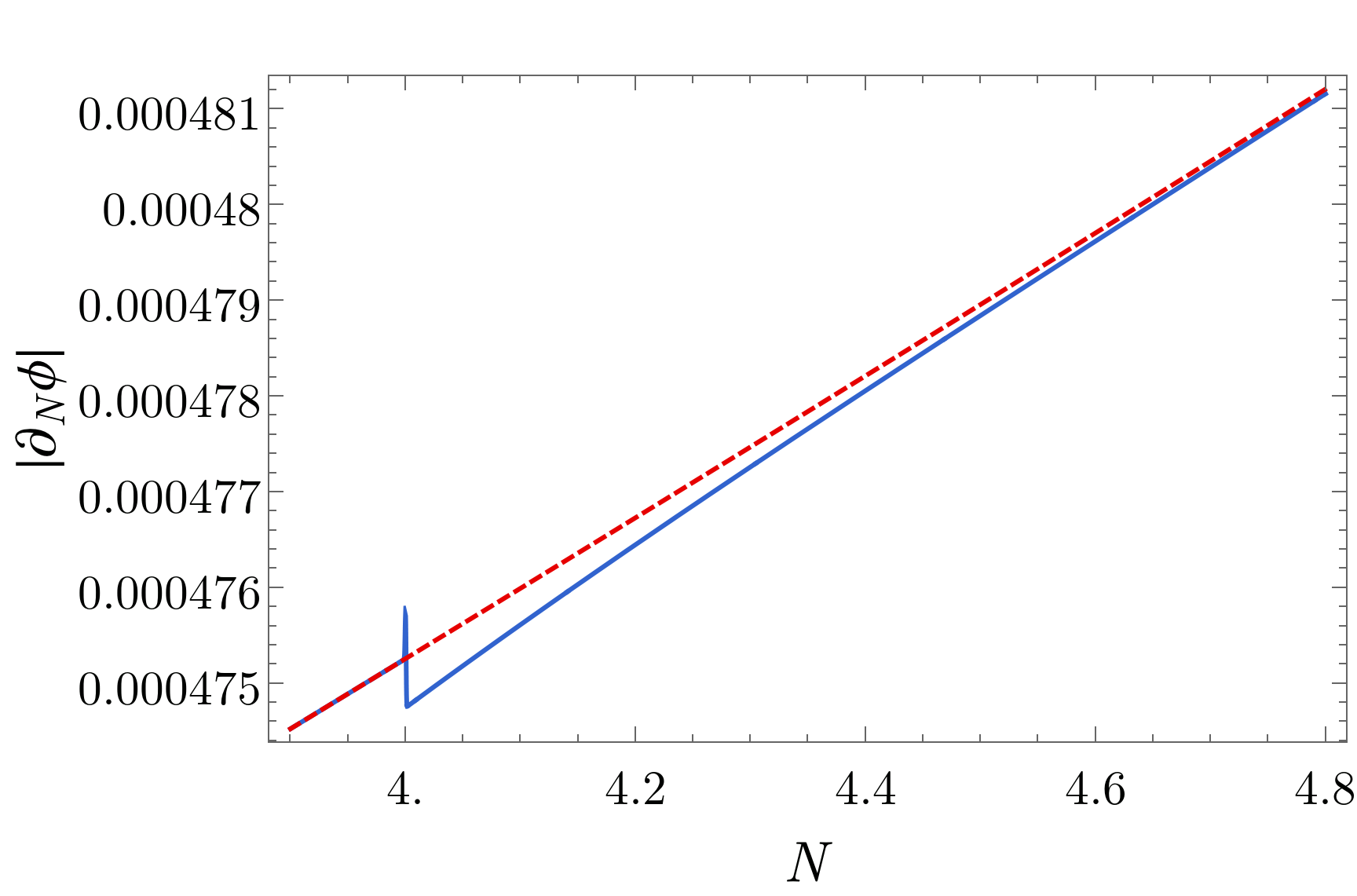}
  \hspace{0.03\textwidth}
  \includegraphics[width=0.4\textwidth]{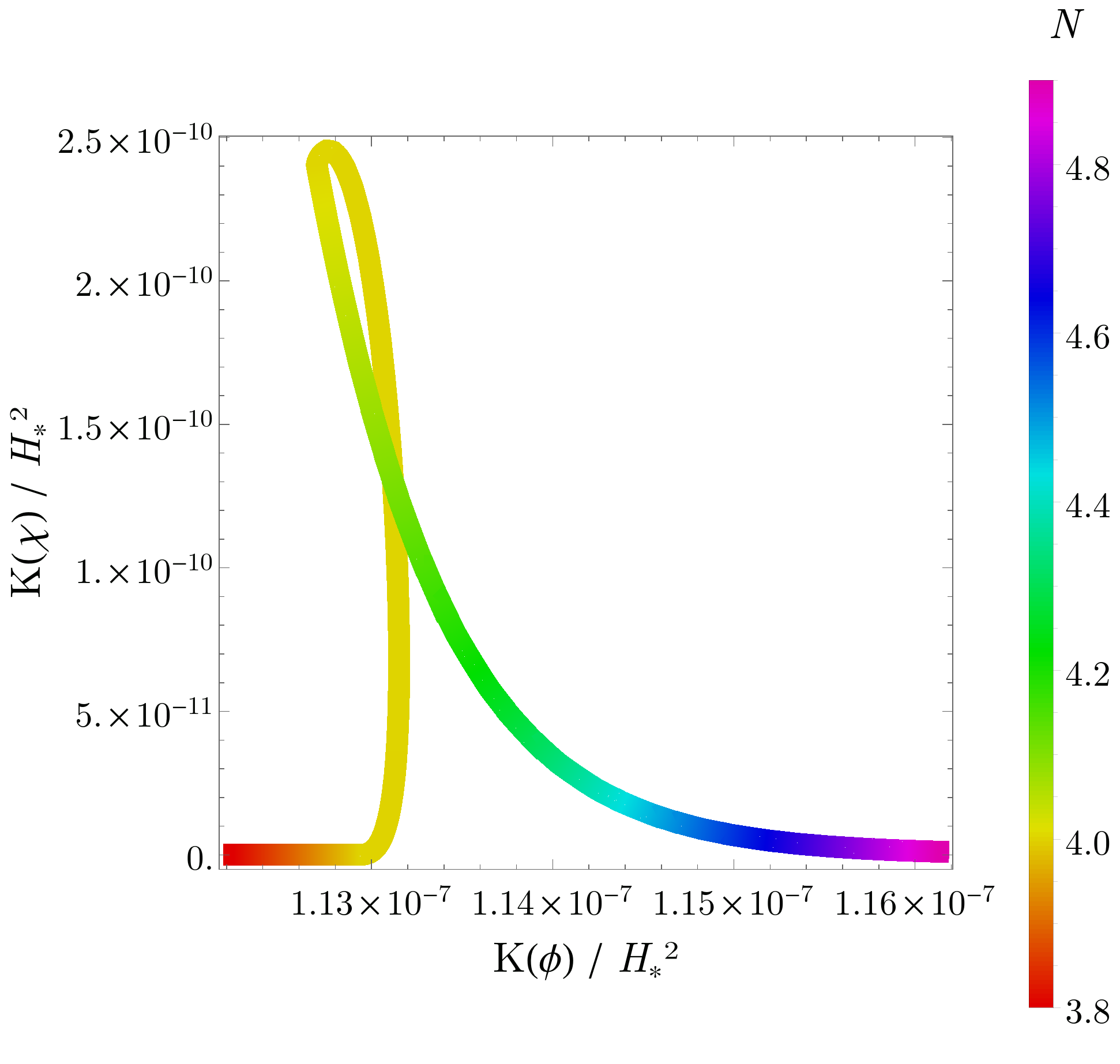}
  \caption{\label{fig:neg_traj} Left panel: $\partial_N \phi$ as a function of $N$. The red dashed line is for without feature scattering, and the blue solid line is with feature scattering. Note that the $\phi$ field loses its kinetic energy very sharply at the 4-th e-fold (where the feature scattering is happening), and the kinetic energy slowly recovers after of order 1 e-fold. Right panel: The kinetic energy $K(\phi)/H_*^2 = (\partial_N \phi)^2$ and $K(\chi)/H_*^2 = (\partial_N \chi)^2$. The color on the plot denotes e-folding number. One finds that about 0.1\% of the $\phi$ field kinetic energy transfers to $\chi$.  Here we have taken $\chi_* = H_* / (2\pi)$ right before hitting the feature.}
\end{figure*}

\begin{figure*}[htbp]
  \centering
  \centerline{
  \includegraphics[width=0.35\textwidth]{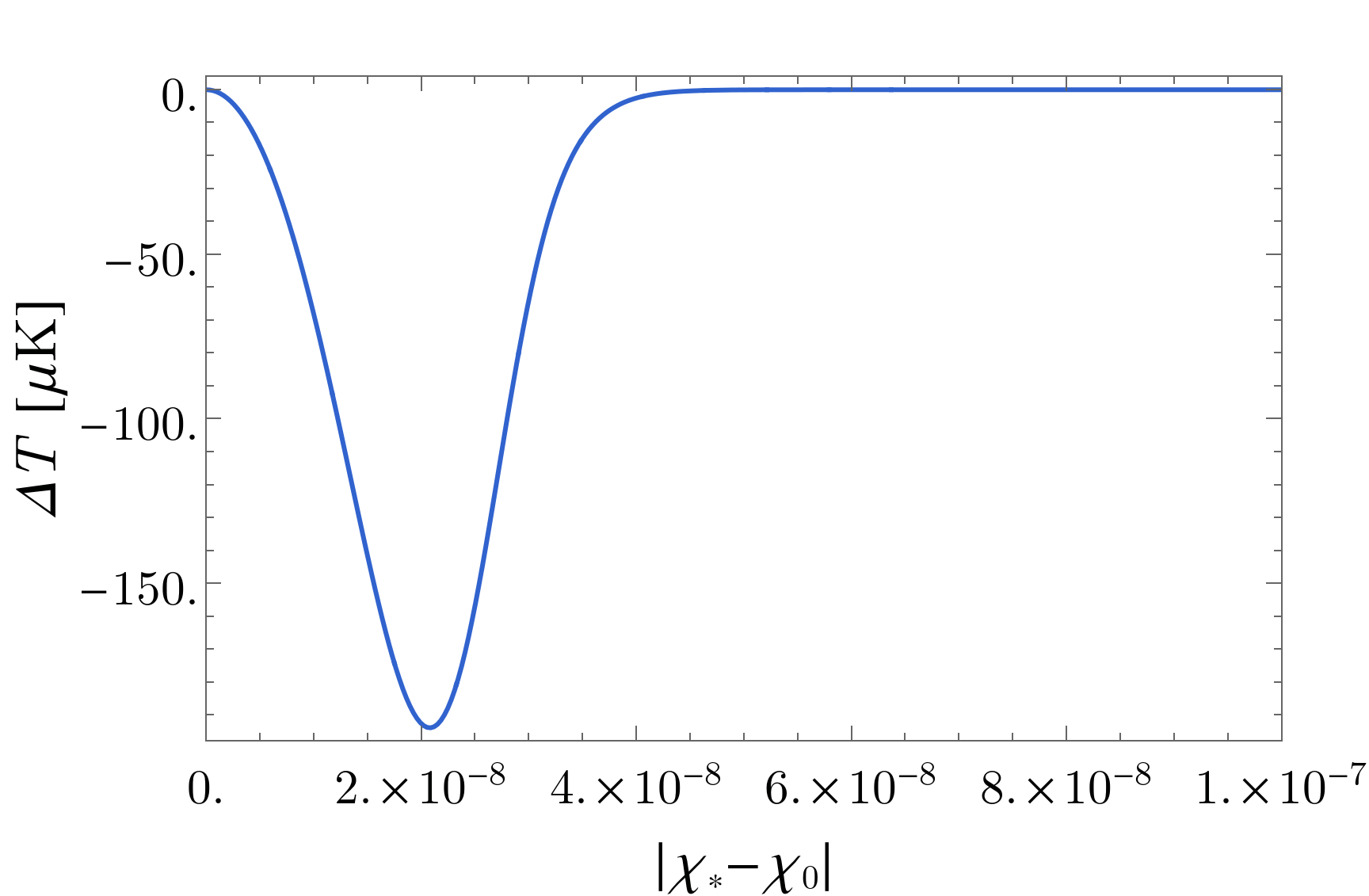}
  \includegraphics[width=0.35\textwidth]{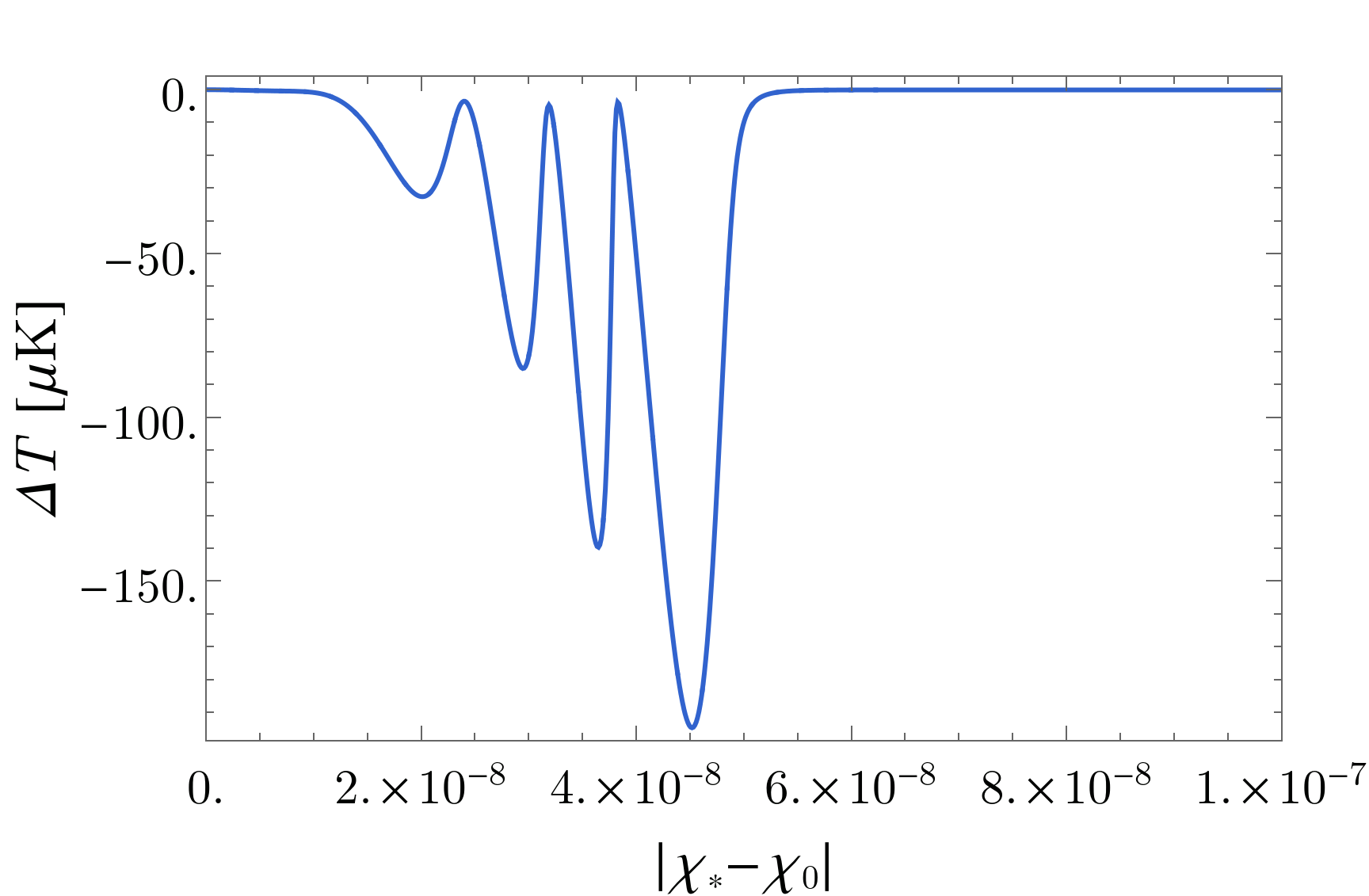}
  \includegraphics[width=0.35\textwidth]{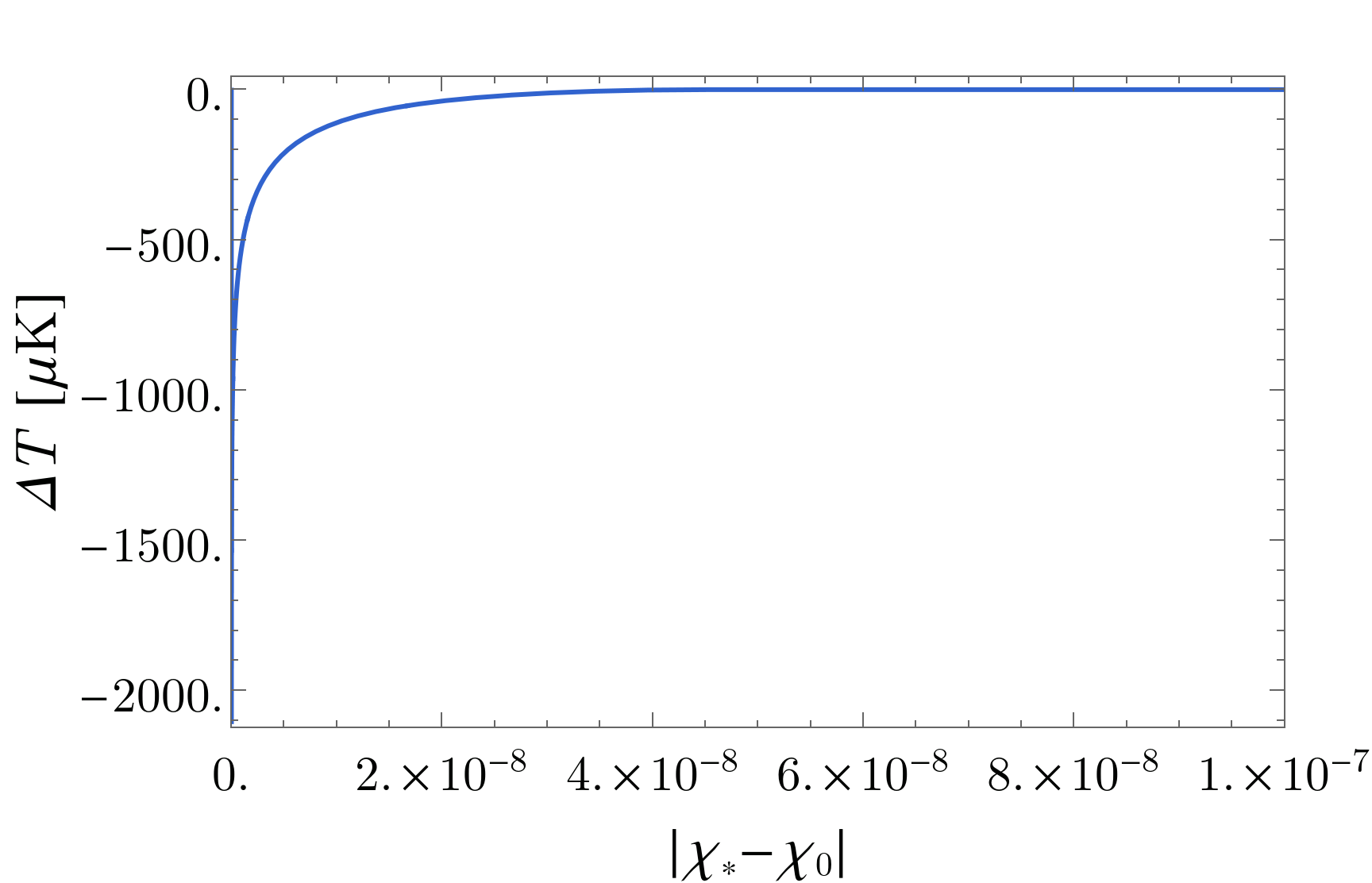}}
  \caption{\label{fig:neg_chi_DT} The temperature fluctuation as a function of $\chi_* - \chi_0$. The left, middle and right panels are for parameters of Benchmark 1,~2,~3 respectively.}
\end{figure*}

\begin{figure*}[htbp]
  \centering
  \centerline{
  \includegraphics[width=0.45\textwidth]{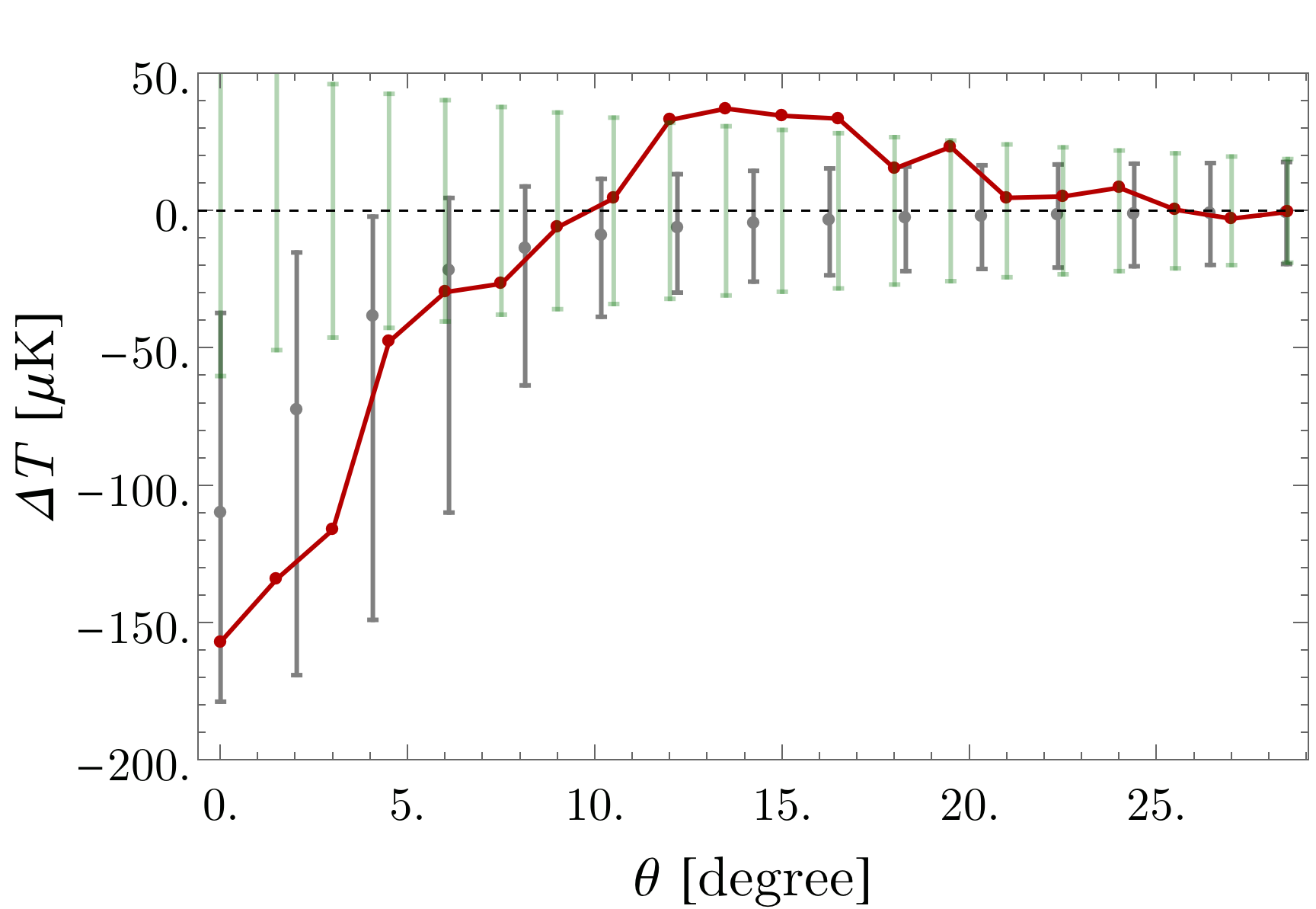}
  \includegraphics[width=0.45\textwidth]{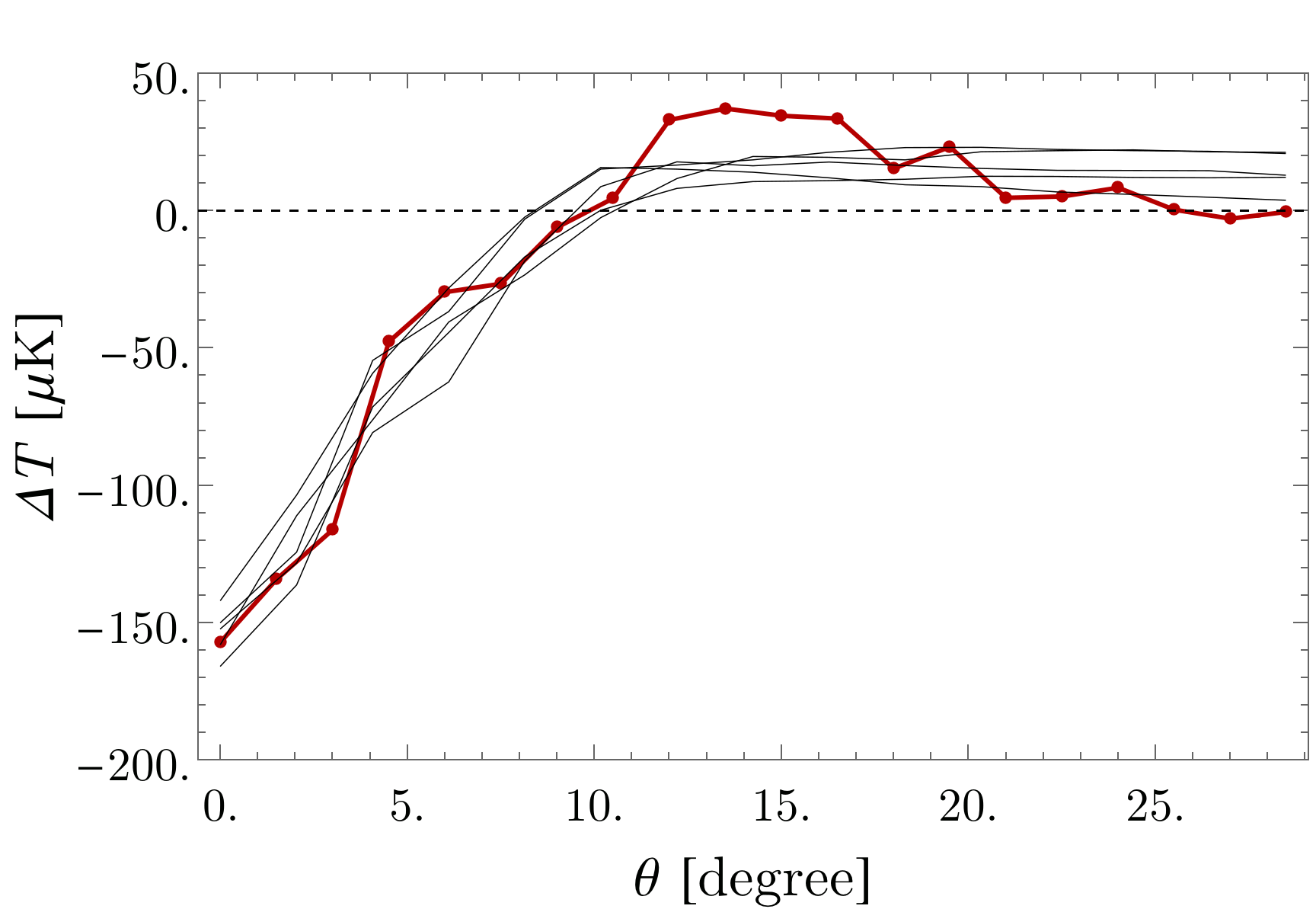}}
 \centerline{\includegraphics[width=0.45\textwidth]{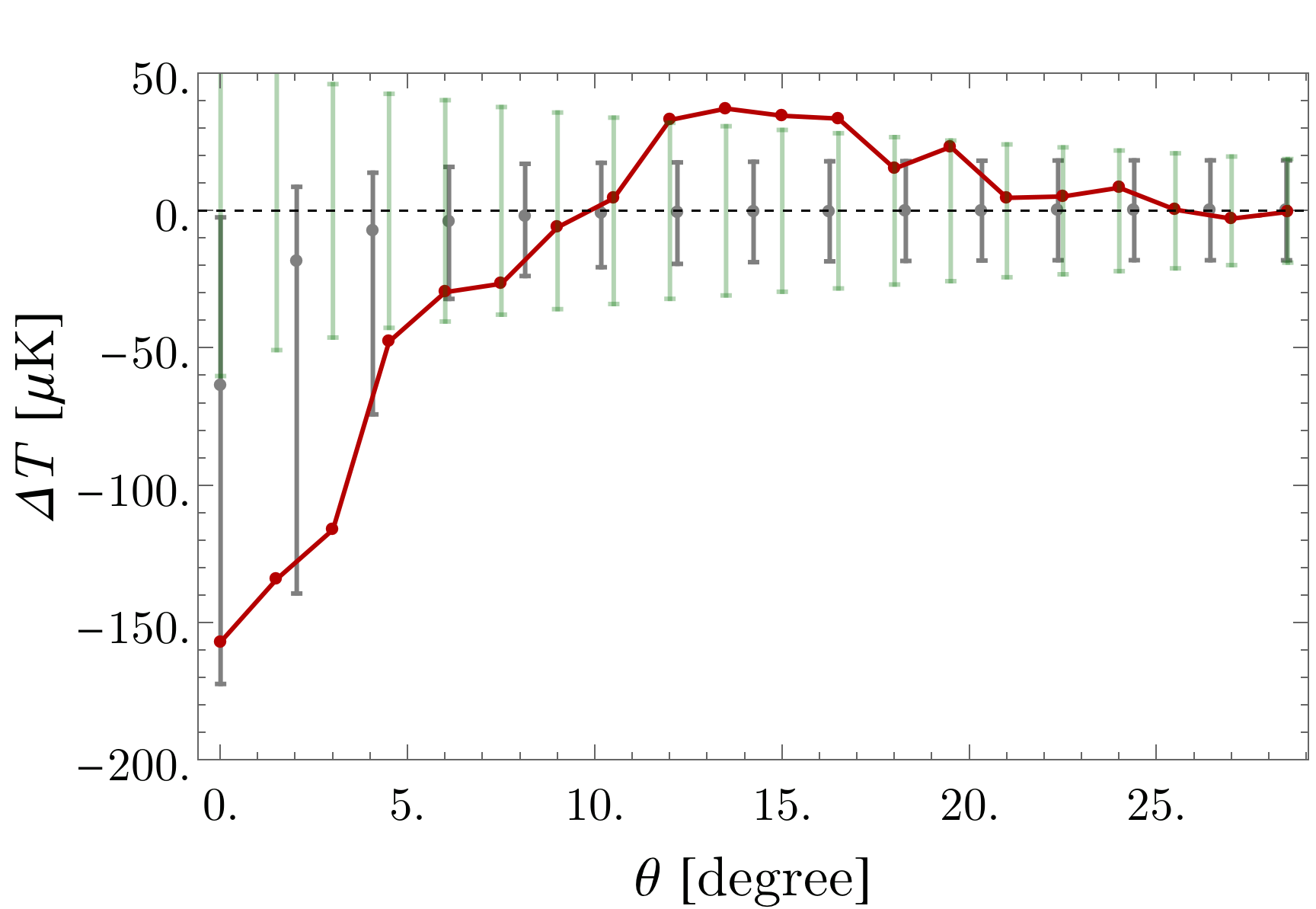}
  \includegraphics[width=0.45\textwidth]{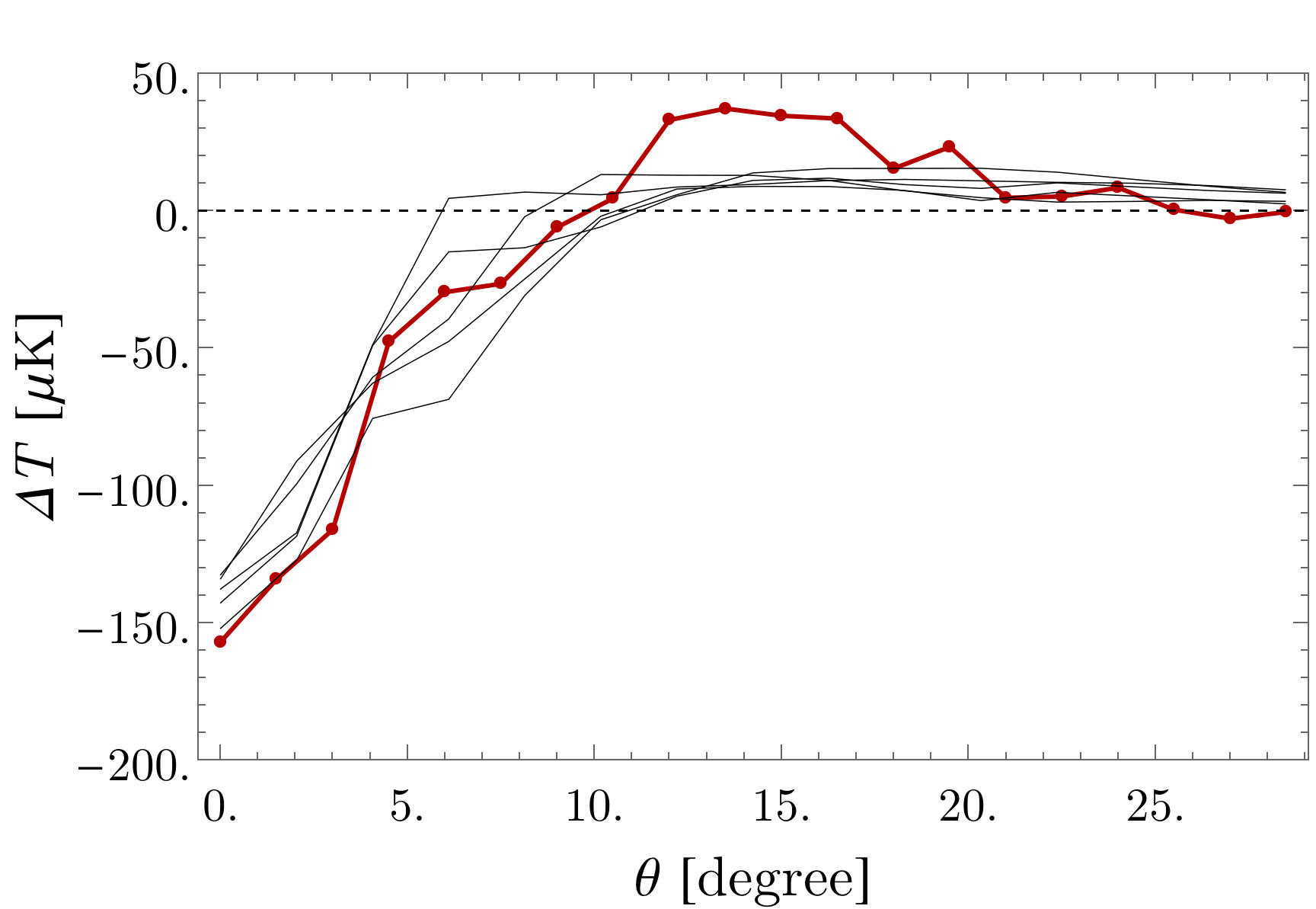}}
\centerline{\includegraphics[width=0.45\textwidth]{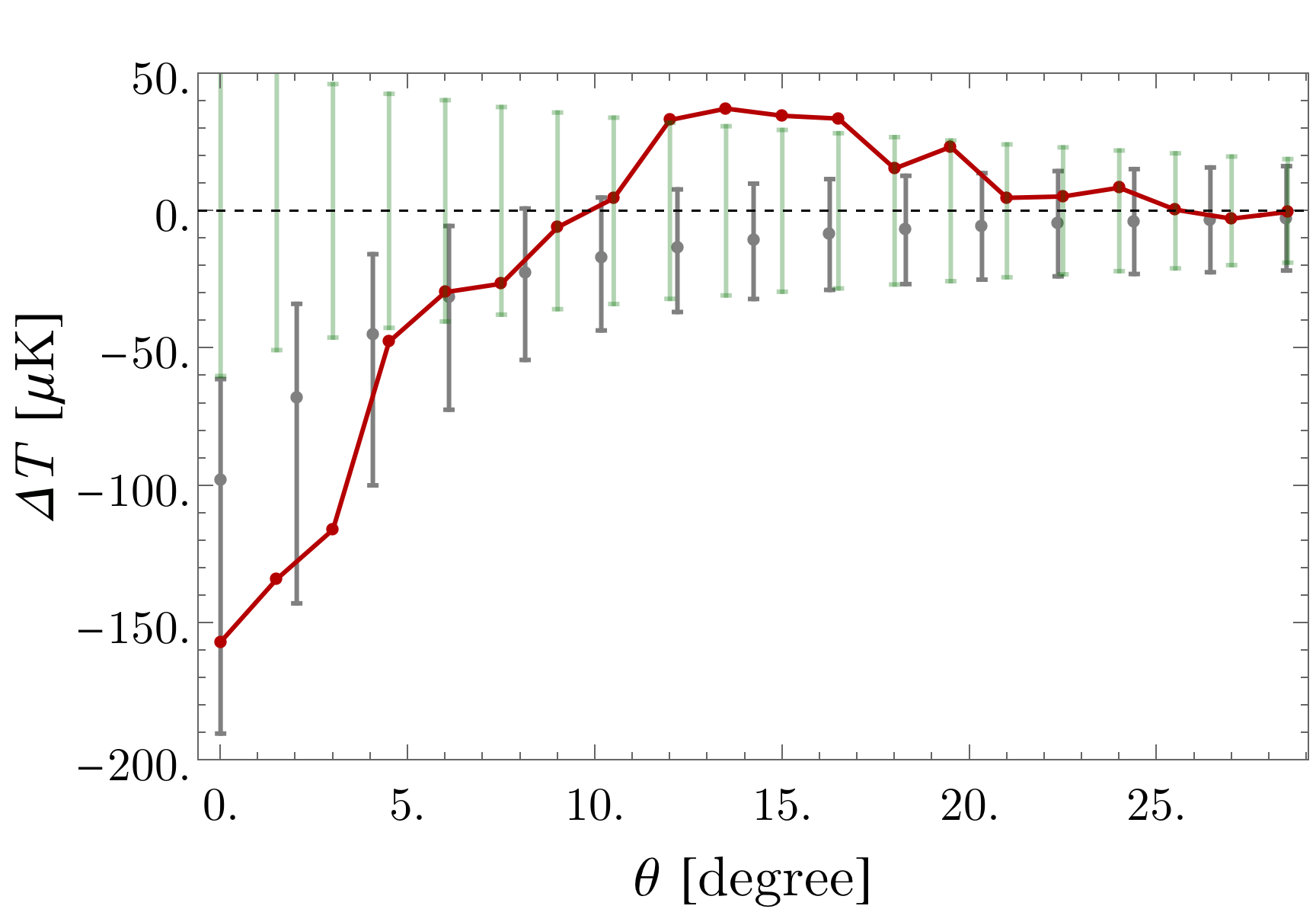}
  \includegraphics[width=0.45\textwidth]{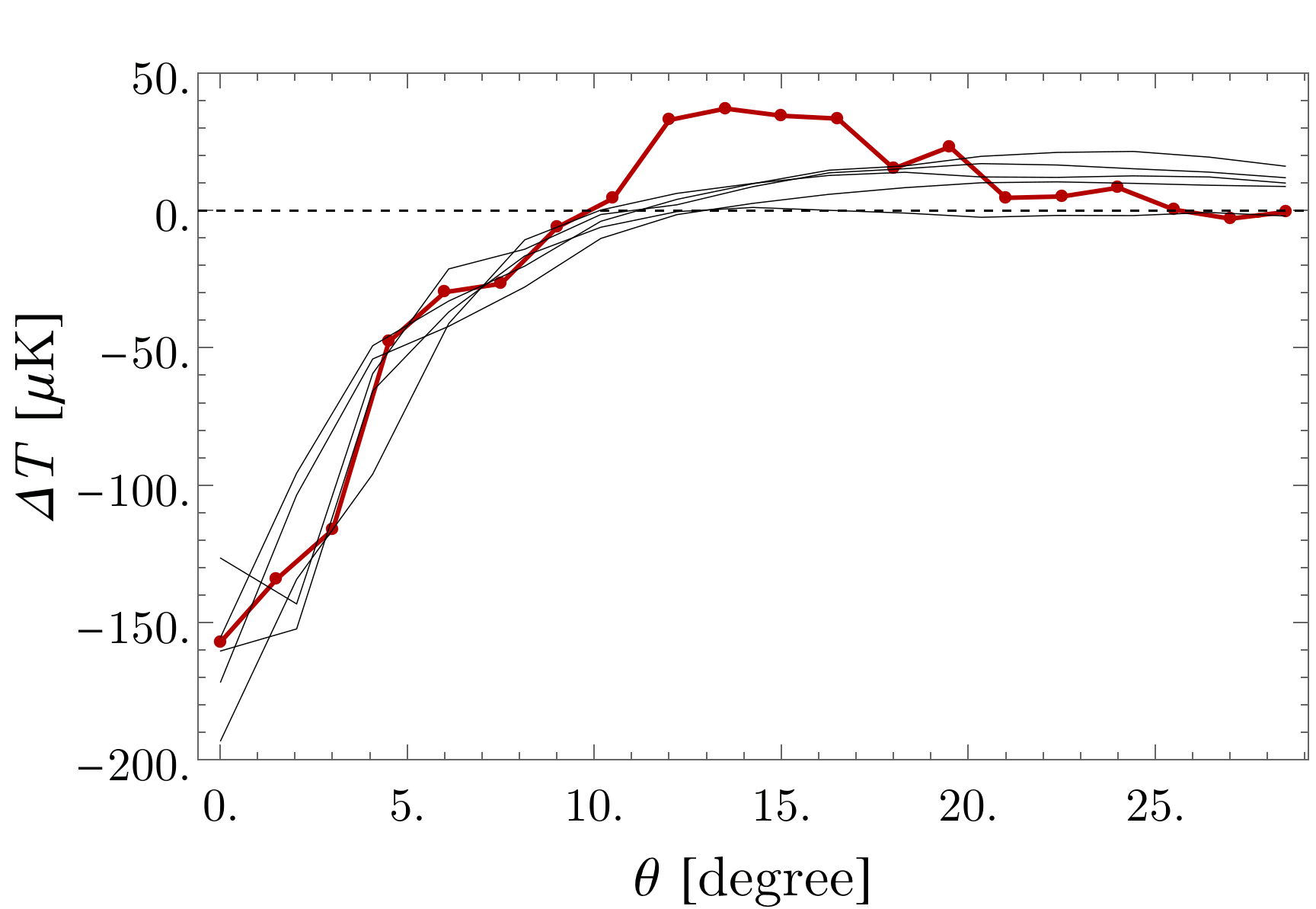}}
  \caption{\label{fig:neg_theta_DT} Left column: The cold spot profile (we have rotated the map such that $\theta=0$ corresponds to the center of the cold spot). The gray bars are 1000 simulations of fluctuations of the feature model. The green bars are simulation of isotropic Gaussian fluctuations (5000 simulated maps). Right column: five best-fitting cold spot profiles from the Benchmark models. The red dot is the actual cold spot from the {\it Planck} {\tt SMICA} map for both panels. The three rows from top to bottom are the Benchmark models 1,~2,~3 respectively. Here the Sachs-Wolfe approximation is used when numerically fitting the bubble profile. The Sachs-Wolfe approximation works well only on large scales. Thus the fitting of the first two points with simulation may not be accurate enough. But one can see that the error bars of those two points from simulation are also large so that the inaccuracy should not affect the physical result significantly.}
\end{figure*}

\begin{figure*}[htbp]
  \centering
 \includegraphics[width=1\textwidth]{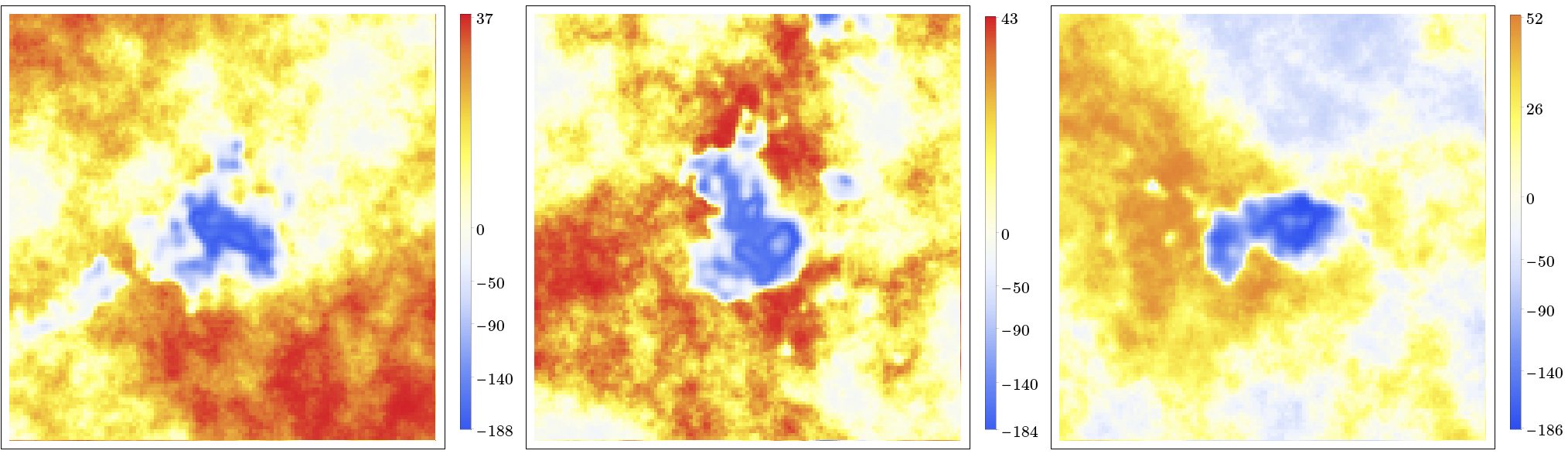}
 \includegraphics[width=1\textwidth]{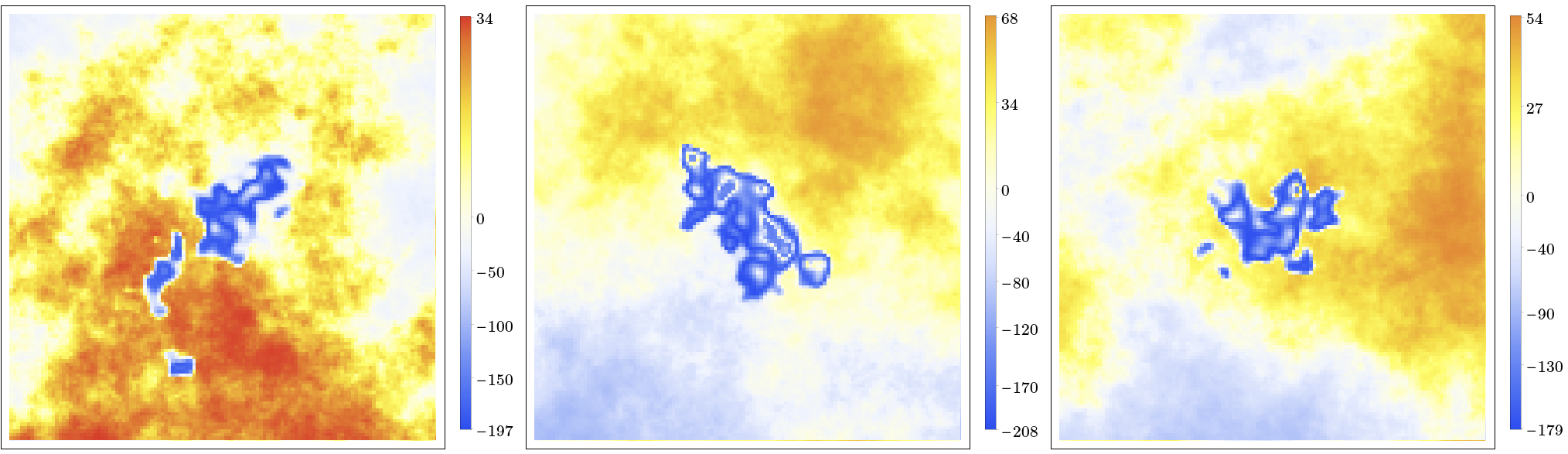}
 \includegraphics[width=\textwidth]{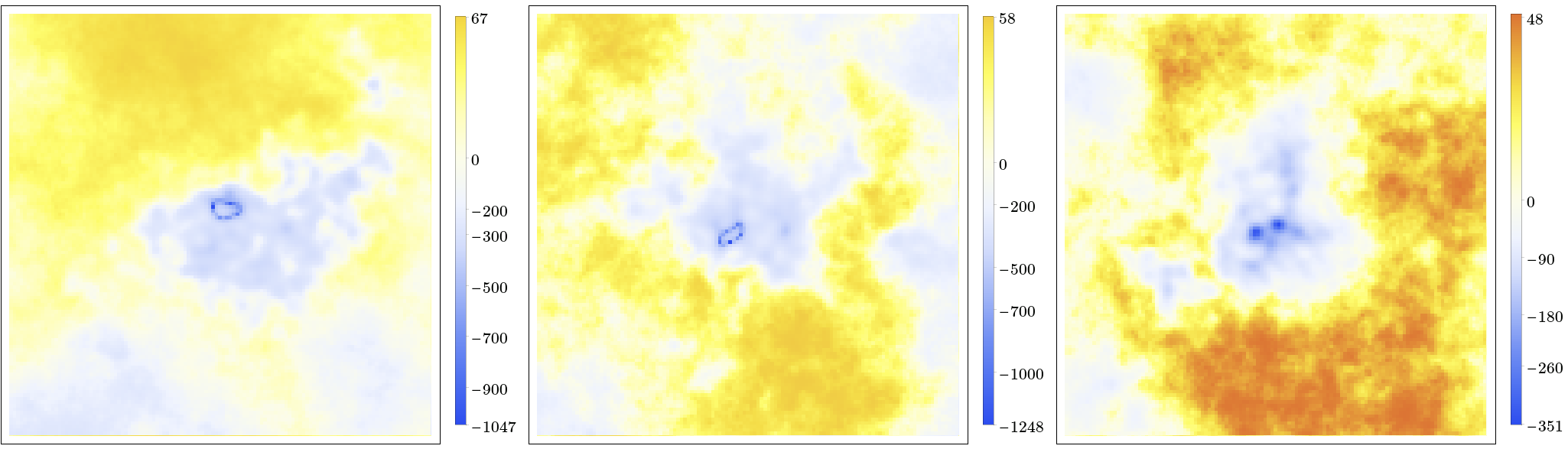}
  \caption{\label{fig:neg_eg} (From top to bottom.) A selection of the best-fitting maps for Benchmark 1,~2~3 respectively. The unit on the plot legend is $\mu$K.}
\end{figure*}





In this section, we provide an explicit example of feature scattering to illustrate the physical mechanism. The generalization to other inflationary potentials should be straightforward. Consider inflation with two fields, both have standard kinetic term, and the potential is given by
\begin{eqnarray}
\label{eq:pot}
  V &=&  V_\mathrm{sr}(\phi) + \delta V~, \nonumber \\
  \delta V &=&  A \exp \left[ -\frac{(\phi-\phi_0)^2}{2\sigma_\phi^2} - \frac{(\chi-\chi_0)^2}{2\sigma_\chi^2}\right] ~.
\end{eqnarray}
The classical initial trajectory is chosen to be on $\phi$ direction, i.e. $\chi=0$.
Here $V_\mathrm{sr}(\phi)$ is the slow-roll part of the potential. Our analysis is not sensitive in the form of $V_\mathrm{sr}(\phi)$, as long as it successfully drives slow-roll inflation. For illustration purpose, we choose a small field potential for $V_\mathrm{sr}(\phi)$:
\begin{align}
  V_\mathrm{sr}(\phi) = V_0 \left( 1 - \frac{\mu^4}{\phi^4} \right)~.
\end{align}
Such a potential can be derived from brane dynamics of string theory \cite{Dvali:1998pa}. We will nevertheless not import anything other than this potential from string theory, but consider it as a phenomenological example.

To fit the observations, we will take $\mu=0.01$ and $V_0 \simeq 4.89 \times 10^{-14}$ \cite{Ma:2013xma}. At the horizon crossing scale of the cold spot, the inflaton field is $\phi_* \simeq 0.153$ for $N_{\ast} \simeq 4$ (counting from the start of the observable stage of inflation), and the Hubble parameter is $H_* \simeq 1.28 \times 10^{-7}$. The feature $\delta V$ is put at $\phi_0 = \phi_*$.

In the isocurvature direction, the initial value of the field $\chi_*$ has no classical preferred value. We set $\langle \chi_* \rangle = 0$ (one can always shift the $\chi_0$ parameter to satisfy this requirement if needed). The quantum fluctuation of $\chi_*(\mathbf{x})$ will be crucial in the analysis. Following the cosmic perturbation theory, $\chi_*(\mathbf{x})$ has a scale invariant power spectrum in Fourier space and we normalize the power spectrum to be $P_\chi = [H/(2\pi)]^2$. In position space, $\chi_*(\mathbf{x})$ behaves as a Gaussian random field. As we will show, the nonlinear mapping from $\chi_*(\mathbf{x})$ to $\Delta T(\mathbf{x})$ generates the cold spot on the CMB.

To explore the parameter space of $\delta V$, we will choose three sets of parameters as three illustrative examples of feature scattering.

\begin{itemize}
\item Benchmark 1: negative short feature. We set $A = -1.1 \times 10^{-5} V_0 \mu^4 / \phi_*^4$, $\sigma_\phi = 10 H_* / (2\pi)$, $\sigma_\chi = 0.7 H_* / (2\pi)$ and $\chi_0 = 4 H_* / (2\pi)$. The procedure of the calculation is also introduced in this subsection.
\item Benchmark 2: negative long feature. We set $A = -1.4 \times 10^{-5} V_0 \mu^4 / \phi_*^4$, $\sigma_\phi = 50 H_* / (2\pi)$, $\sigma_\chi = 0.7 H_* / (2\pi)$ and $\chi_0 = 5.2 H_* / (2\pi)$. The differences from Benchmark 1 are discussed.
\item Benchmark 3: positive feature. We set $A = 1.0 \times 10^{-4} V_0 \mu^4 / \phi_*^4$, $\sigma_\phi = 50 H_* / (2\pi)$, $\sigma_\chi = 0.5 H_* / (2\pi)$ and $\chi_0 = 3.5 H_* / (2\pi)$. The differences from Benchmarks 1 and 2 are discussed.
\end{itemize}

Note that in the above benchmarks, the randomness in the potential needs not to be large. In fact, $\delta V / V \sim 10^{-5}$ is already significant enough to bring observable difference. This is because, the isocurvature fluctuation control whether the inflation trajectory hit the feature. And once the trajectory hits the feature, the change of curvature perturbation will be order $\zeta \sim \delta V / V \sim 10^{-5}$. As a result, our mechanism is extremely sensitive to small features in the potential.

The $\delta N$ formalism \cite{Starobinsky:1986fxa, Sasaki:1995aw, Lyth:2004gb} will be used to investigate the cosmological perturbations from the feature scattering. The $\delta N$ formalism uses the observation that different Hubble patches during inflation can be approximated as different local FRW universes. Thus the cosmological observables can be calculated by exploring those local FRW universes. Especially, the curvature perturbation in the uniform energy density slice can be calculated by \footnote{See the appendix of \cite{Chen:2007gd} for clarification of sign conventions.}
\begin{align}
  \zeta =  \delta N~,
\end{align}
where $\delta N$ is the difference of e-folding number between an initially-flat slice and a finally-uniform energy density slice. The curvature perturbation $\zeta$ is then converted to the CMB temperature anisotropy following the standard theory of CMB. In this paper we will use the Sachs-Wolfe approximation $\delta T/T \simeq - \zeta/5 = - \delta N/5$. Intuitively, the relation can be understood as:
\begin{align}
&   \mbox{Inflaton scattered by feature}  \nonumber \\
& \rightarrow  \mbox{smaller kinetic energy in $\phi$ direction} \nonumber \\
& \rightarrow \mbox{ larger $\delta N$}   \nonumber \\
 & \rightarrow \mbox{later reheating} \nonumber \\
 &  \rightarrow \mbox{less dilution of energy} \nonumber
\\
& \rightarrow \mbox{higher energy density} \nonumber \\
&  \rightarrow \mbox{deeper gravitational potential} \nonumber \\
& \rightarrow \mbox{lower CMB temperature}~.
\end{align}
More precise relation between $\delta T/T$ and $\zeta$ can be studied by solving the Boltzmann equations at recombination.

Note that $\delta N$ has two sources: the quantum fluctuation of $\phi$ and the quantum fluctuation of $\chi$. Those independent sources can be studied independently, and they needs to be added together to make the total curvature fluctuation. In the numerical calculation, we will focus on investigating $\delta N$ as a function of $\chi$ but we will add back a contribution of Gaussian random field to account for the contribution from $\phi$ at the last map-making stage.

The angular size of the cold spot is around 10 degrees. Converting to horizon crossing time of comoving wave number, this corresponds to $\ell \sim 20$. Note that in the observed CMB temperature power spectrum, there is indeed a dip at $\ell \sim 20$. It will be interesting to do a combined analysis on both signals, from feature scattering.

In the following subsections, we will explore different parameter space of Eq.~(\eqref{eq:pot}) with three benchmarks, and calculate the e-folding number as functions of the position of isocurvature field $\delta N = \delta N(\chi)$.

\subsection{Benchmark 1: Negative short feature}

In this subsection, we consider a negative short feature in the potential. We set $A = -1.1 \times 10^{-5} V_0 \mu^4 / \phi_*^4$, $\sigma_\phi = 10 H_* / (2\pi)$, $\sigma_\chi = 0.7 H_* / (2\pi)$ and $\chi_0 = 4 H_* / (2\pi)$. The size of the feature in the $\phi$ direction is relatively small (compared to the case of Benchmark 2). As a result, the $\sigma$ field does not have enough time to oscillate inside the negative feature.

Before getting to the implications to the late universe, it is intuitive to check what was happening during inflation.

%

As an example, we consider the $\chi$ field value when hitting the feature to be $\chi_* - \chi_0 = H_*/(2\pi)$. The time evolution of $\partial_N \phi$ is plotted on the left panel of Fig.~\ref{fig:neg_traj}. One can find from the plot that, at $N_* \simeq 4$ (where feature scattering happens)
\begin{align}
  \left| \frac{{\rm d}\phi}{{\rm d} N} \right| \sim 4 \times  10^{-4}~.
\end{align}
When the inflationary trajectory hits the feature, there is a change in the velocity of the inflaton $\phi$. At first, $|{\rm d}\phi/{\rm d}N|$ increase because $\phi$ falls into a potential well. However, two other effects immediately follow: First, the obtained energy is returned because $\phi$ has to climb out of the potential well; second, the potential well can be considered as a scattering center, such that a similar amount of energy is transferred to the $\chi$ direction (as evidenced in the right panel of Fig.~\ref{fig:neg_traj}). As a result, there is a overall loss of velocity of the inflaton, with the value read from the plot
\begin{align}
  \delta \left| \frac{{\rm d}\phi}{{\rm d}N} \right| \sim - 4 \times  10^{-7} \sim
  - 10^{-3} \left| \frac{{\rm d}\phi}{{\rm d}N} \right|~.
\end{align}
In words, the inflaton field has lost $0.1\%$ of its kinetic energy because of hitting the feature in the potential \footnote{Here we do not need to consider the change of potential energy before/after the feature scattering, because we only need to compare the change of energy before and after a very sharp scattering process, and the change of potential energy is negligible.}. Note that the inflaton always loses its kinetic energy because of scattering off a feature. Thus a cold spot instead of a hot spot is predicted \footnote{When the potential is positive, the inflaton actually gains kinetic energy at the very beginning, when it falls into the potential well. However, the gained energy is returned after a time interval that is much smaller than Hubble time. Thus the effect of gaining the kinetic energy is negligible when we integrate to obtain the e-folding number. On the other hand, the loss of kinetic energy through feature scattering needs order of 1 e-fold to recover, and thus it is the dominate effect.}.

Note that the loss of kinetic energy takes of order one e-fold to recover, because this is the time scale to reach the inflationary attractor solution. To be more precise, as shown in the left panel of Fig.~\ref{fig:neg_traj}, it takes about roughly $0.2$ e-fold for the inflaton to recover its kinetic energy. As a result \footnote{The $\delta N$ here is from the quantum fluctuation of $\chi$. The part from $\phi$ will be added at the map-making stage.},

\begin{align}
  \delta N = \frac{\delta\phi}{\left| \frac{{\rm d}\phi}{{\rm d}N} \right| + \delta \left| \frac{{\rm d}\phi}{{\rm d}N} \right|}
  = \frac{0.2 \times \left| \frac{{\rm d}\phi}{{\rm d}N} \right|}
  {\left| \frac{{\rm d}\phi}{{\rm d}N} \right| + \delta \left| \frac{{\rm d}\phi}{{\rm d}N} \right|}
  \sim 2 \times 10^{-4}~.
\end{align}
From the Sachs-Wolfe approximation, the temperature fluctuation $\delta T/T \simeq -\zeta/5 = -\delta N/5$. Thus we get $\delta T\simeq -100 \mu$K.

Carrying on the above analysis for general values of $\chi_* - \chi_0$, numerically the temperature fluctuation as a function of $\chi_* - \chi_0$ is plotted in the left panel of Fig.~\ref{fig:neg_chi_DT}. We then simulate 1000 maps and rotate those maps such that they center at the cold spot. In the left panel of the top row of Fig.~\ref{fig:neg_theta_DT}, we bin the pixels of the map as a function of $\theta$, and then plot the 0.5, 0.16 and 0.84 quantiles (corresponding to the central value and 1$\sigma$ uncertainty if the probability distribution were Gaussian) of the binned pixel temperature. In the right panel of the top row of Fig.~\ref{fig:neg_theta_DT}, we plot the top five best-fitting cold spot profiles. Three best-fitting examples are given in the top row of Fig.~\ref{fig:neg_eg}.

\subsection{Benchmark 2: Negative long feature}

In this subsection we consider a negative long feature. Inside a ``long'' feature, there is enough time for the $\chi$ field to oscillate. We set $A = -1.4 \times 10^{-5} V_0 \mu^4 / \phi_*^4$, $\sigma_\phi = 50 H_* / (2\pi)$, $\sigma_\chi = 0.7 H_* / (2\pi)$ and $\chi_0 = 5.2 H_* / (2\pi)$. Note that the feature size in the $\phi$ direction is 5 times longer than that of Benchmark 1.

\begin{figure*}[htbp]
  \centering
  \includegraphics[width=0.55\textwidth]{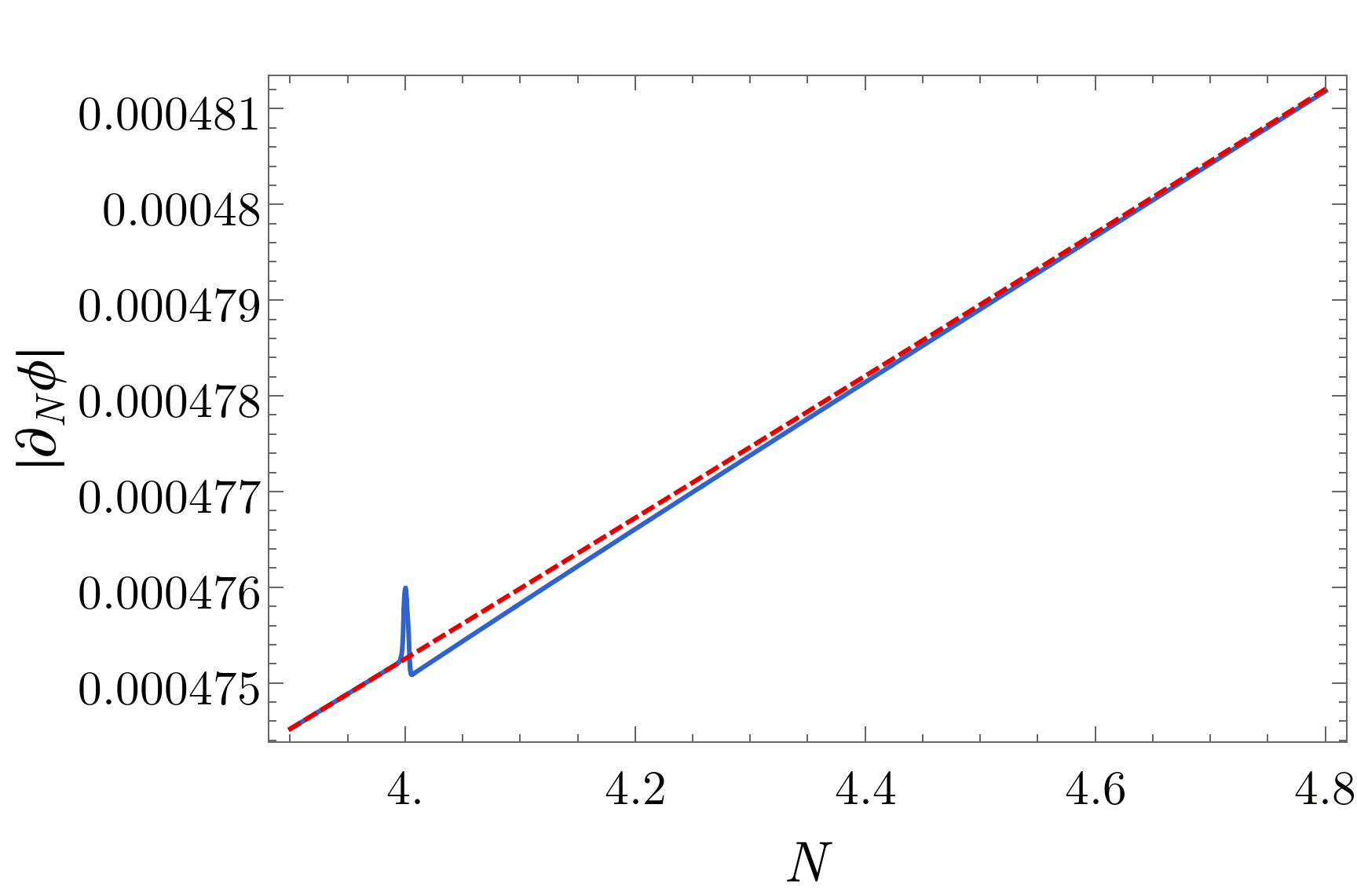}
  \hspace{0.03\textwidth}
  \includegraphics[width=0.4\textwidth]{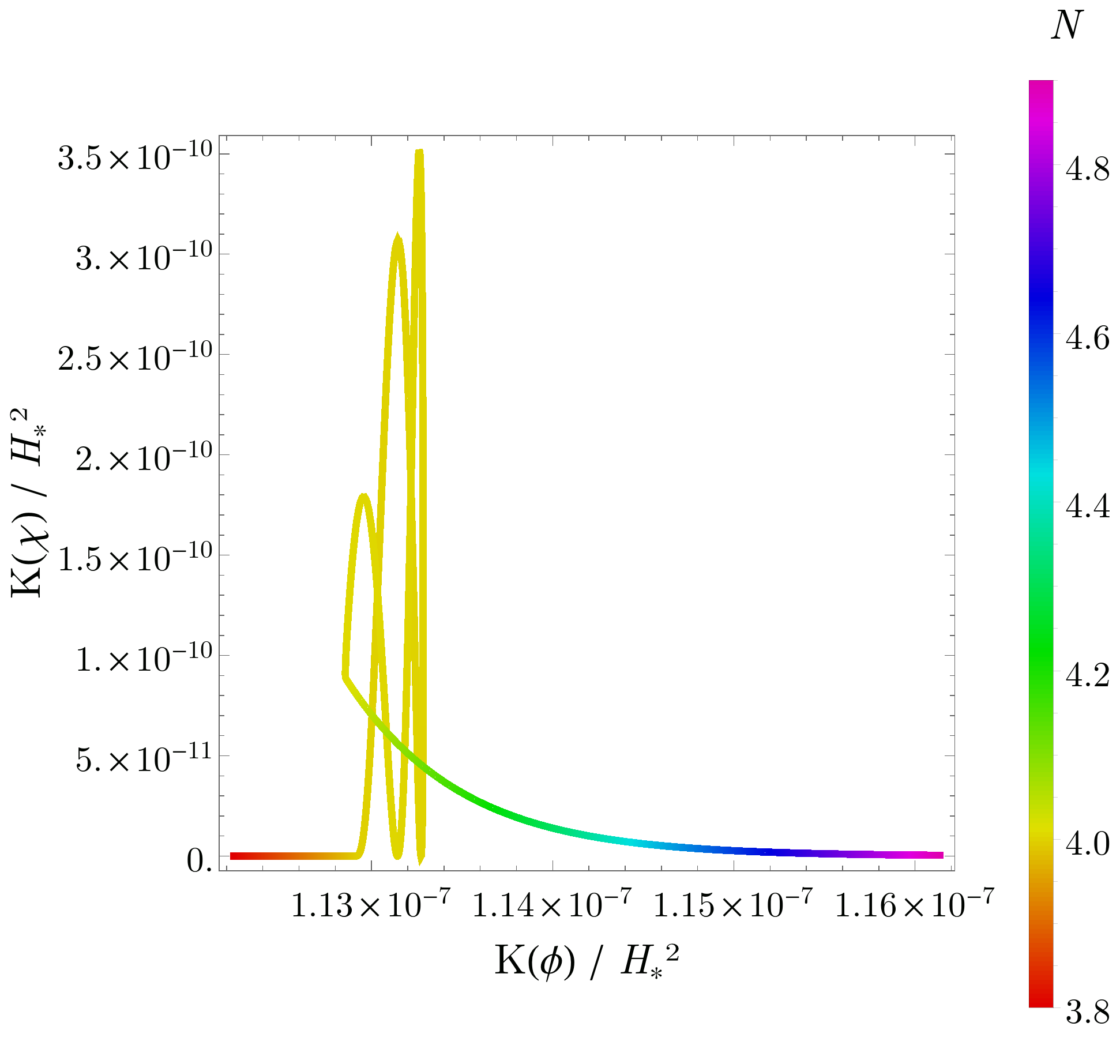}
  \caption{\label{fig:osc_traj} The same as Fig.~\ref{fig:neg_traj} but for Benchmark model 2. Note that the feature in the inflationary potential is long enough to hold a few oscillations in this case.}
\end{figure*}


Finer structures develop in the case of long feature. If we have chosen an initial value $\chi_* - \chi_0 = H_*/(2\pi)$, we find similar behavior to Fig.~\ref{fig:neg_traj}. However, for some other initial values, for example,  $\chi_* - \chi_0 = 1.4 H_*/(2\pi)$, the $\chi$ field oscillates inside the feature. The situation is illustrated in Fig.~\ref{fig:osc_traj}. As a result, some amount of $\chi$ kinetic energy is returned to $\phi$. For some special values, almost all kinetic energy is returned. Thus oscillatory patterns develop, as shown in the middle panel of Fig.~\ref{fig:neg_chi_DT}.

The oscillations in $\delta T(\chi)$ leads to ring objects in $\delta T(\mathbf{x})$. The cold spot in this parameter space is not completely cold, but instead has nested rings of cold and hot patterns. There is no evidence of such patterns in the actual CMB cold spot. There has been a debate if there are other ring patterns in the sky \cite{Gurzadyan:2011ac}. The Benchmark 2 parameters provide an explanation for such rings (though the shape is not perfectly spherical). However, it is shown that the appearance of such rings is due to an inappropriately chosen power spectrum, and thus not actually in the sky \cite{Eriksen:2011fi}. Thus we will not tune the parameters to fit such features here.

The general and best-fitting cold-spot profiles, and sample spots are plotted in the middle row of Figs.~\ref{fig:neg_theta_DT} and \ref{fig:neg_eg}, respectively.

\subsection{Benchmark 3: Positive feature}

In this subsection we consider positive feature scattering. We set $A = 1.0 \times 10^{-4} V_0 \mu^4 / \phi_*^4$, $\sigma_\phi = 50 H_* / (2\pi)$, $\sigma_\chi = 0.5 H_* / (2\pi)$ and $\chi_0 = 3.5 H_* / (2\pi)$. With positive potential, the $\chi$ direction has a run-away solution and can never be bounded. So there is no oscillation in this case.

The $\delta T(\chi)$ dependence is not as sharp as the previous two benchmarks. Thus the cold spot does not have as clear a boundary as those previous cases. Because of the not-so-sharp $\delta T(\chi)$ dependence, to get a cold enough center of the spot, we have to increase $A$. As a result, the cold spots generated by Benchmark 3 typically (though not always) have very cold tinny cores.

The general and best-fitting cold-spot profiles, and sample spots are plotted in the bottom row of Figs.~\ref{fig:neg_theta_DT} and \ref{fig:neg_eg}, respectively.

Considering a very sharp and cold core is formed for positive features, it is unlikely that such features can fit the realistic cold spot. But considering that the sharpness of the cold core depends on model parameters. The cold core may appear at $\ell > 2000$ scales and free streaming may erase the sharpness.

\section{Massive isocurvature directions}
\label{sec:mass-isoc-direct}

\begin{figure*}[htbp]
  \centering
  \includegraphics[width=0.45\textwidth]{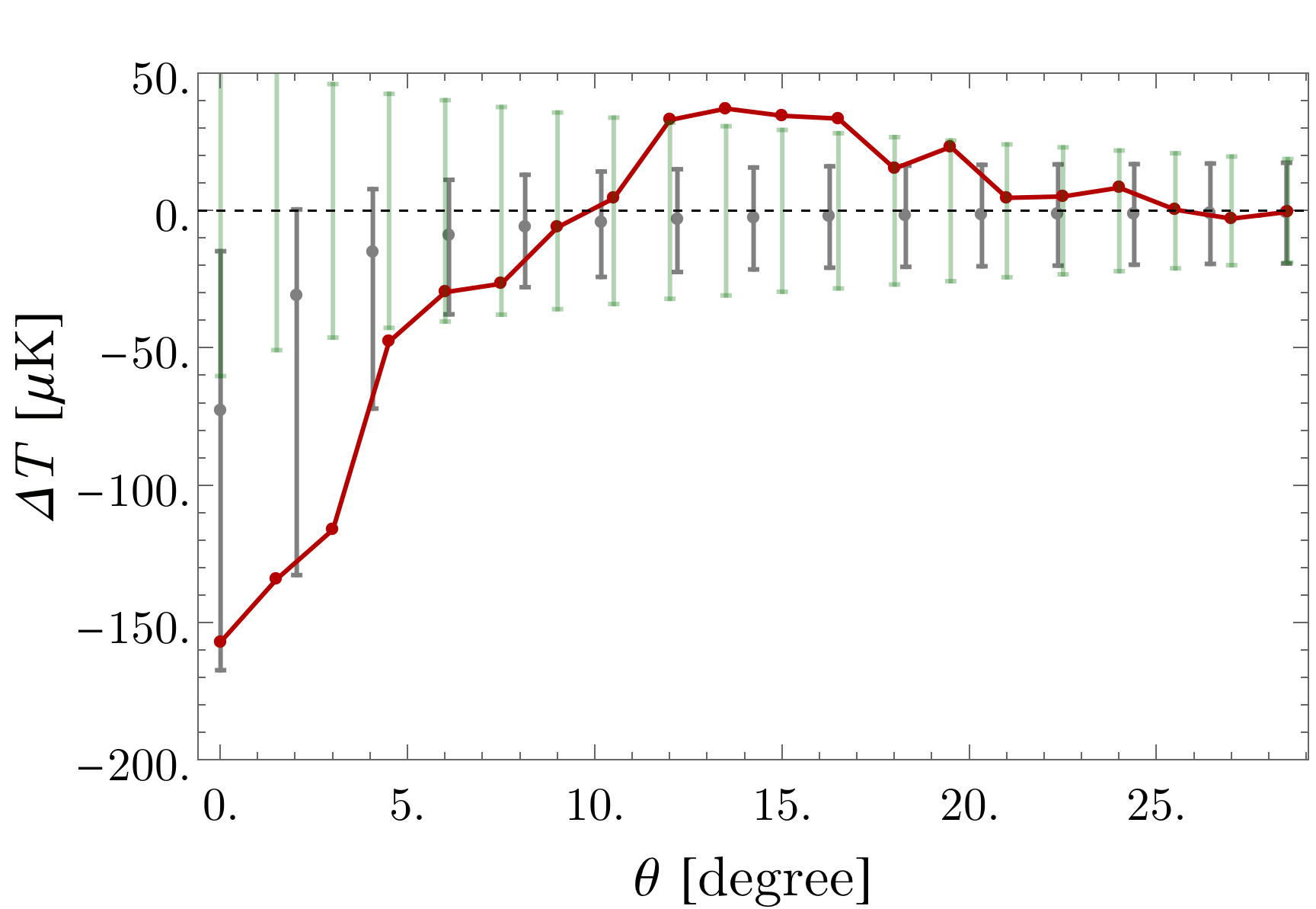}
  \includegraphics[width=0.45\textwidth]{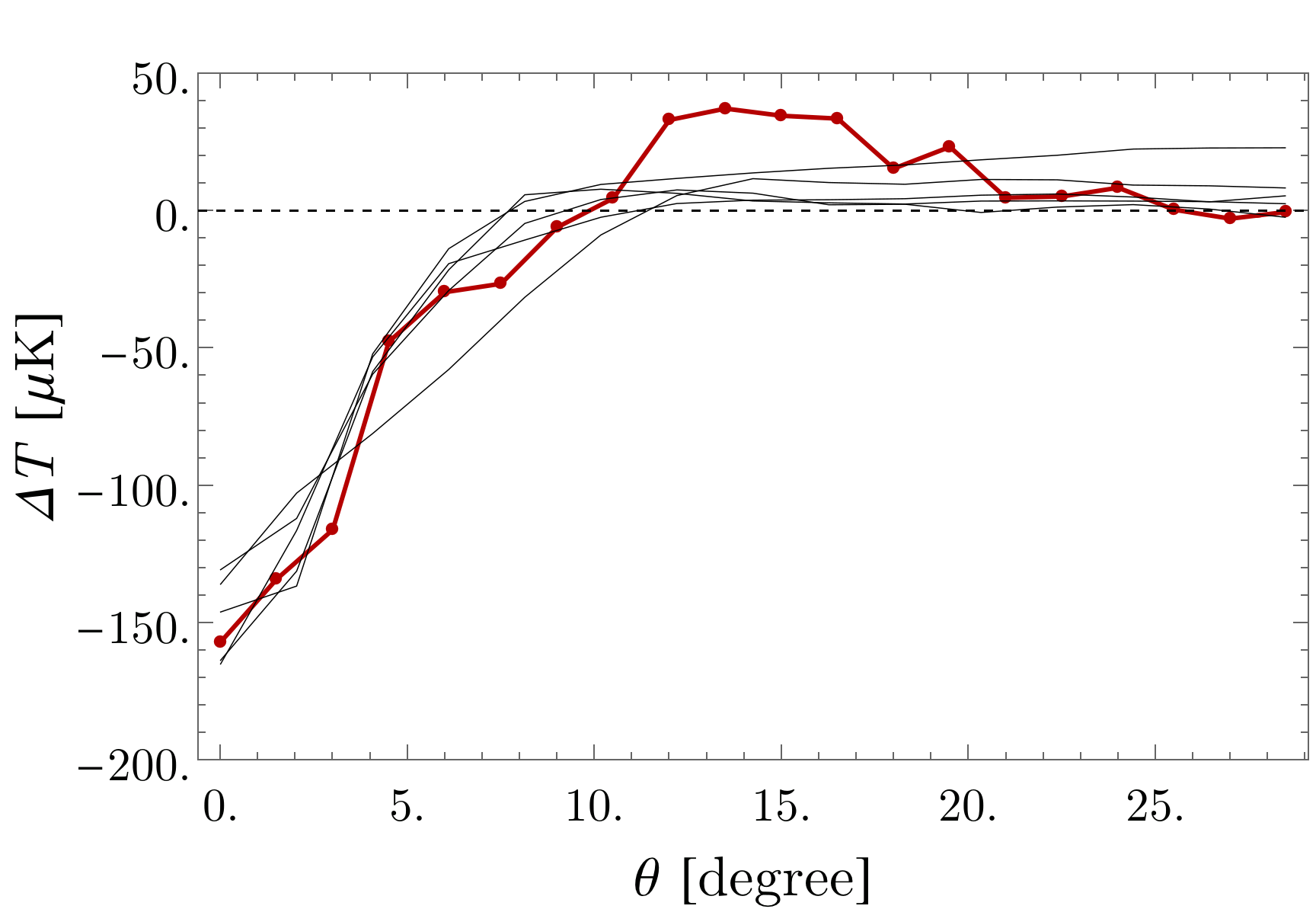}  \caption{\label{fig:mas_theta_DT} The same as Fig.~\ref{fig:neg_theta_DT} but for massive isocurvature model $m_{\chi}=0.9H_{\ast}$.}
\end{figure*}

\begin{figure*}[htbp]
  \centering
  \includegraphics[width=\textwidth]{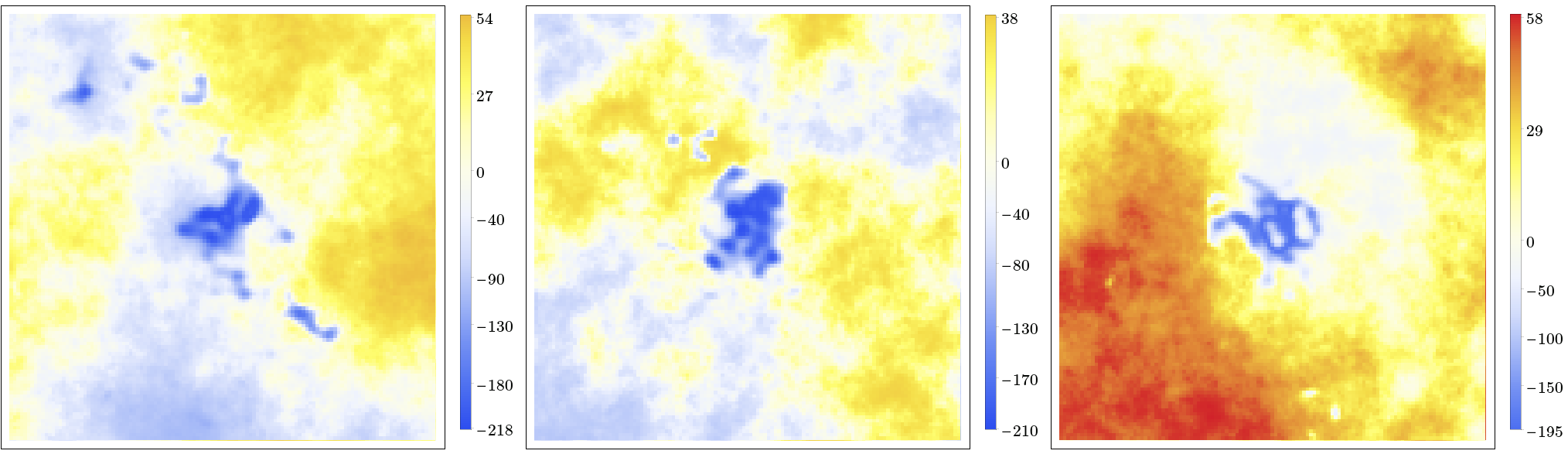}
  \caption{\label{fig:mas_eg} The same as Fig.~\ref{fig:neg_eg} but for model $m_\chi = 0.9 H_*$.}
\end{figure*}

The feature scattering not only applies to massless isocurvature directions, but also to marginally massive ones. As long as the mass of the isocurvature direction is not much greater than $H$, the energy scale for inflationary fluctuations, the isocurvature direction can still be excited to explore its field space. During inflation, fields with $m_\chi \sim H$ arise naturally \cite{Chen:2009we, Chen:2009zp, Baumann:2011nk, Chen:2012ge}. Thus it is worthwhile to investigate feature scattering of such marginally massive fields, with an additional mass term in the potential
\begin{align}
  V_\mathrm{mass}(\chi) = \frac{1}{2} m_\chi^2 \chi^2~.
\end{align}

Note that the process of feature scattering happens on a time scale much shorter than Hubble. Thus as long as $m_\chi$ is not much great than $H_*$, $m_\chi$ does not enter the dynamics of feature scattering. However,
the spectral index of a massive field is significantly different from a scale invariant spectrum, with spectral index
\begin{align}
n_\chi-1 = \mathrm{Re} \left(  \frac{3}{2} - \sqrt{\frac{9}{4}-\frac{m^2}{H^2}}  \right)~.
\end{align}
Thus the mass of the field controls how many e-folds can the isocurvature direction randomly travel. For example, when $m = 0.9 H_*$, the $\chi$ field has a spectrum $n_\chi - 1 = 0.3$.

In the numerical example, we set $m = 0.9 H_*$, $A = -1.1 \times 10^{-5} V_0 \mu^4 / \phi_*^4$, $\sigma_\phi = 10 H_* / (2\pi)$, $\sigma_\chi = 0.7 H_* / (2\pi)$ and $\chi_0 = 4.4 H_* / (2\pi)$. In other words, we have only tuned $\chi_0$ to be slightly greater, and consider a massive $\chi$ field. Otherwise the parameters coincide with the Benchmark 1 of massless case.

The general and best-fitting cold-spot profiles, and sample spots are plotted in Figs.~\ref{fig:mas_theta_DT} and \ref{fig:mas_eg}, respectively.

As one can see from Fig.~\ref{fig:mas_eg}, there are more small-scale structures compared to the case of massless Benchmark 1.

\section{Conclusion and discussion}
\label{sec:concl-disc}
To conclude, we propose a scattering mechanism of inflaton due to the features in the inflation potential, during which the isocurvature fluctuations are converted into curvature fluctuations. The curvature fluctuations direction loses kinetic energy due to the scatters in isocurvature direction, therefore the number of e-folds becomes larger in some region of the universe. We find that the cold spot can be well explained by such a mechanism, and the spot profile reasonably fits the CMB cold spot without fine tuning of the inflationary parameters. Before ending up the paper, we would like to mention a few future directions:

\begin{itemize}

\item Beyond the Sachs-Wolfe approximation. Currently we did not use the full Boltzmann code to calculate the CMB transfer function. The Sachs-Wolfe approximation works well for exploring the coarse-grained cold-spot profile but is not enough for probing the fine structures inside the spot. One can use the full radiation transfer function to carry on a more detailed analysis in the future.

\item Verifying the approximations of the $\delta N$ formalism. The $\delta N$ formalism assumes that the horizon-crossing amplitude of the $\phi$ and $\chi$ quantum fluctuations are Gaussian and uncorrelated. In the case of feature scattering, this is not rigorously true because of the turning of trajectory. However, note that the transfer of the inflaton kinetic energy is tinny (0.1\% in the studied example), the correction from sub-horizon physics should be suppressed by this small fraction.

Having that said, it remains interesting to see if the calculation from $\delta N$ formalism can be verified by first principle calculation. However, such a calculation is challenging. The first principle calculation of cosmological perturbations is known as the in-in formalism. But in the in-in formalism, the primary calculatables are the correlation functions. In other words, one starts from the Gaussian two point correlation function and studies small departure from that. But the cold spot is highly non-Gaussian and localized object, and thus is not easily captured by the correlation functions from the in-in formalism.

\item Extra species/symmetry point (ESP) \cite{Kofman:2004yc, Battefeld:2011yj} as scattering centers. In our current examples, the inflaton kinetic energy is lost into the collective motion of the isocurvature direction. It may also be possible that the kinetic energy of the inflaton is lost into hidden ESPs in the isocurvature direction. It is interesting to explore such possibilities.

\item Connection between the feature scattering regime and the multi-stream regime: By carefully arranging the position of the feature, it is possible that different part of the universe follows different classical trajectories, separated by temporary domain walls. This is known as the multi-stream inflation \cite{Li:2009sp, Afshordi:2010wn}, which is also a possible explanation of the cold spot. Feature scattering and multi-stream corresponds to different regimes of the multi-field parameter space. If the features on the inflationary potential are random, we expect the feature scattering to be more typical (whereas highly random features in multi-stream inflation cause disasters and provide a constraint on multi-field inflation \cite{Duplessis:2012nb}).

\item Connection between the cold spot and the dip of CMB temperature power spectrum at $\ell \sim 20$. We argue they may come from the same origin -- feature scattering during inflation. With our current toy potential given at Eq.~(\eqref{eq:pot}), we are unable to explain this power deficit at the same time. However, a more general shaped potential may be able to explain both the deficit and the cold spot.

\item String theory model building. It is believed that string theory has a landscape of complicated vacuum structures \cite{Bousso:2000xa, Susskind:2003kw}. It would be interesting to build string landscape models for feature scattering. Especially, our mechanism assumes that $\sigma_\phi$ and $\sigma_\chi$ have value of order $H$. This assumption is made at the phenomenological level in this paper. Nevertheless, it is interesting to search for theoretical motivations for such a potential. For inflation to work, the inflationary perturbations, in the sense of effective field theory, should have enough range of validity. Especially, the UV cutoff of the theory should be at least of order $H$. In the marginal case where the UV cutoff of the effective field theory is just of order $H$, new features are expected when the field rolls a distance of order $H$. In addition, inspired by quasi-single field inflation \cite{Chen:2009we, Chen:2009zp}, the mass parameters during inflation should be naturally of order $H$. Thus once there are many mass parameters in the landscape, such as some mass matrix, variations will be introduced in the potential with field range of order $H$. We leave detailed investigation of those possibilities to future works.

\item Careful model comparison between our feature scattering mechanism and the $\Lambda$CDM base model. We have 5 parameters in our illustrative model. The parameters are chosen to be the convenient ones for inflation model building, but not optimized to fit the data as a phenomenological model. It remains interesting to propose a more economical model with fewer number of parameters, and perform model comparison between our mechanism and standard inflation model.

\end{itemize}

\section*{Acknowledgments}
We would like to thank the helpful discussion with Richard Battye and Clive Dickinson. YW was supported by Grant HKUST4/CRF/13G issued by the Research Grants Council (RGC) of Hong Kong, a Starting Grant of the European Research Council (ERC STG grant 279617), and the Stephen Hawking Advanced Fellowship. YZM acknowledges support from an ERC Starting Grant (no.~307209).

\appendix

\section{Generation of Gaussian random fields}


\begin{figure}[htbp]
  \centering
  \includegraphics[width=0.45\textwidth]{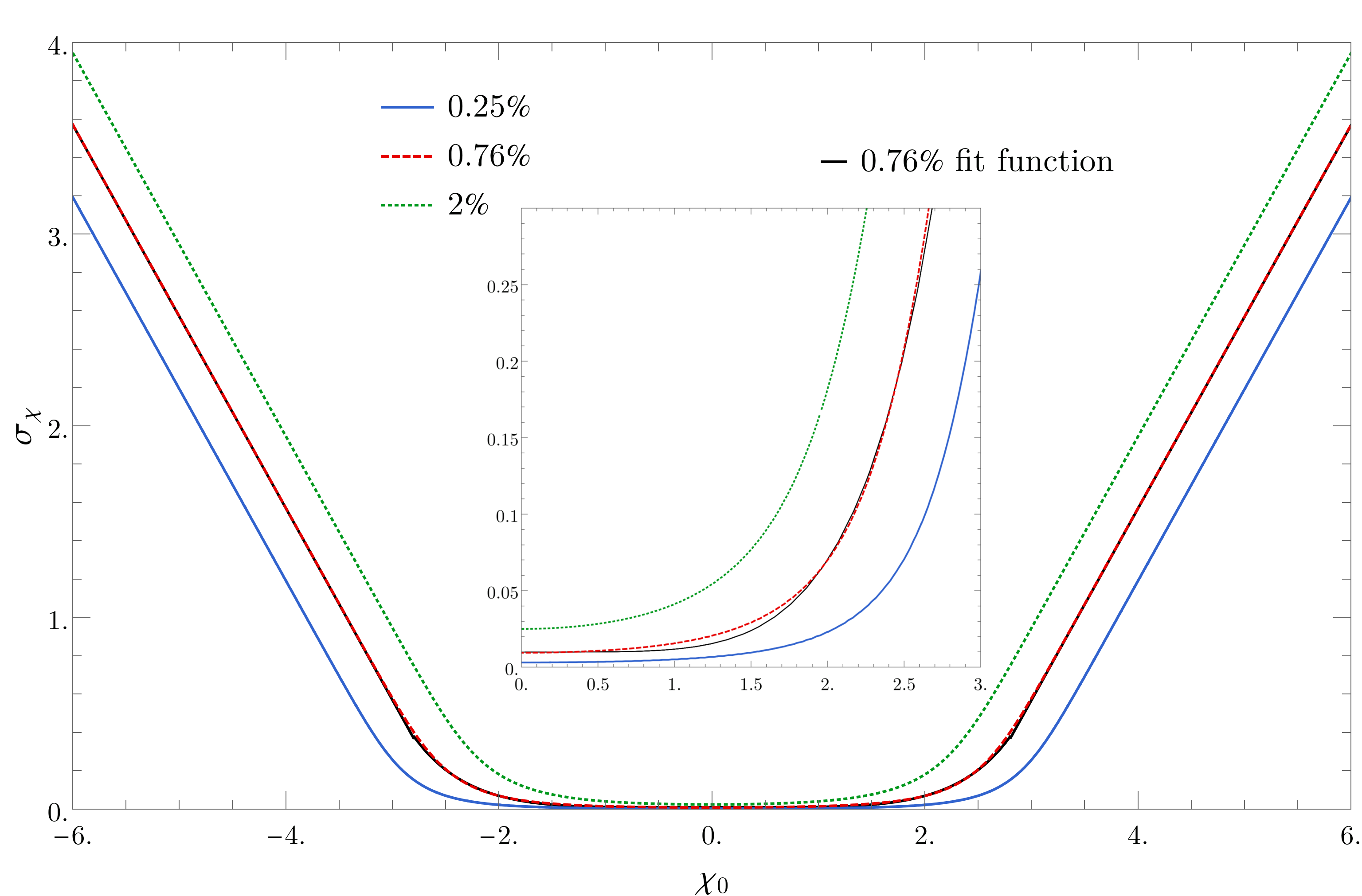}
  \caption{\label{fig:fitChoice} Relation between $\chi_0$ and $\sigma_\chi$, in order to keep the cold spot(s) covering 0.76\% of the sky. The $\chi_0$ axis is plotted with unit $c=H_* / (2\pi)$. }
\end{figure}

\begin{figure*}[htbp]
  \centering
  \includegraphics[width=1\textwidth]{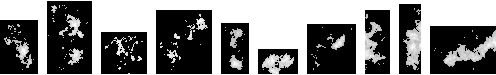}
  \includegraphics[width=1\textwidth]{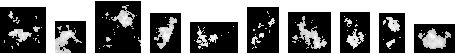}
  \caption{\label{fig:figure_label} Samples of rare fluctuations out of a Gaussian random field. The upper and lower panels show samples of $2.5\sigma$ and $3\sigma$ rare fluctuations respectively.}
\end{figure*}

The fluctuation of the isocurvature direction follows Gaussian statistics. In this appendix, we provide some details of statistics.

The cold spot has an area of about 0.76\% of the full CMB sky. To have such a fraction of sky covered, there are two possibilities: Either the feature is very narrow ($\sigma_\chi \ll H_*$), or $\chi_0$ is about $2.5\sigma$ away from $\sigma_\chi$.
Practically, the former tends to produce a number of small cold spots and the latter tends to produce one or a few large cold spot (which is the case of our current interest).

To be more explicit, the relation between $\sigma_\chi$ and $\chi_0$ is
\begin{align}\label{eq:chiRelation}
  \int_{-\sigma_\chi}^{\sigma_\chi} \frac{1}{c\sqrt{2\pi}} e^{-\frac{(\chi-\chi_0)^2}{2c^2}} {\rm d}\chi \sim 0.0076~,
\end{align}
where $c \equiv \sqrt{\langle\chi_*^2\rangle}$. Here we have used ``$\sim$'' because there are uncertainties from both different realizations of random variables, and the detailed profile of the inflationary potential (considering the feature does not sharply start at $-\sigma_\chi$ and sharply end at $\sigma_\chi$). Equation \eqref{eq:chiRelation} does not have an analytic solution. The numerical solution is plotted in Fig.~\ref{fig:fitChoice}, and can be fitted to high precision by
\begin{align} \sigma_\chi =
\begin{cases}
  \chi_0-2.43c, & \text{if } \chi_0>2.8c \\
  -\chi_0-2.43c, & \text{if } \chi_0<-2.8c \\
  \left[\exp \left( \frac{|\chi_0/c|^5}{546} \right) -0.99\right] c, & \text{if } -2.8c \leq \chi_0 \leq 2.8c
\end{cases} ~. \label{eq:fitRelChoice}
\end{align}
This relation is plotted and referred to as ``0.76\% fit function'' in Fig.~\ref{fig:fitChoice}.
In the numerical examples, $\sigma_\chi$ and $\chi_0$ are chosen around Eq.~\eqref{eq:fitRelChoice}, and slightly adjusted considering the detailed profile of the potential.

\section{Selection of spots}

In the numerical simulation of the CMB maps, the cold spot is selected with the following pipeline:
\begin{itemize}
\item The simulated sky is a 200$\times$200 square degrees. This approximately resembles the whole CMB sphere. Note that we will eventually focus on the spot, which is around 10$\times$10 square degrees, thus the flat sky approximation can be applied here for technical simplicity.
\item For each map, we generate a mask, where the unmasked pixels have temperature $\delta T < - 100 \mu$K.
\item The mask is smoothed with a Gaussian filter with a width of 2 degrees. After Gaussian smoothing, the very nearby ones are connected.
\item The largest connected (after smoothing) spot on the mask is selected.
\item We crop the real image according to the spot position of the mask, with a radius of 30 degree. We drop the images if the spot is too close to the boundary of the whole image.
\end{itemize}

Some examples of spot shapes are illustrated in Fig.~\ref{fig:figure_label}.


\begin{thebibliography}{999}

\bibitem{Hinshaw13} G.~Hinshaw et al., 2013, ApJS, 208, 19

\bibitem{Planck16} Planck Collaboration, Planck 2013 results. XVI, 2014, A\& A, 571, 16

\bibitem{Planck23} Planck Collaboration, Planck 2013 results. XXIII, 2014, A\& A, 571, 23

\bibitem{Vielva04} P.~Vielva et al., 2004, ApJ, 609, 22

\bibitem{Cruz07} M.~Cruz et al., 2007, ApJ, 655, 11

\bibitem{Gurzadyan14} V.~G.~Gurzadyan et al., 2014, A\& A, 566, 135

\bibitem{Chiang06} L.-Y.~Chiang, P.~D.~Naselsky, 2006, Int.~J.~Mod.~Phys.~D 15, 1283-1298

\bibitem{Tojeiro06} R.~Tojeiro et al, 2006, MNRAS, 365, 265

\bibitem{Cruz06} M.~Cruz, M.~Tucci, E.~Martinez-Gonzalez, \& P.~Vielva, 2006, MNRAS, 369, 57–67

\bibitem{Inoue06} K.~T.~Inoue, \& J.~Silk, 2006, ApJ, 648,
23

\bibitem{Rees68} M.~J.~Rees, \& D.~M.~Sciama, 1968, Nature, 217, 511–516.

\bibitem{Finelli14} F.~Finelli et al., arXiv:~1405.1555

\bibitem{Szapudi14} I.~Szapudi et al., 2015, MNRAS, 450, 288

\bibitem{Kovacs14} A.~Kovacs, \& I.~Szapudi, 2014, arXiv: 1401.0156

\bibitem{Nadathur14} S.~Nadathur, M.~Lavinto, S.~Hotchkiss, \& Syksy Rasanen, 2014, Phys.~Rev.~D~90,~103510

\bibitem{Zibin14} J.~P.~Zibin, 2014, arXiv: 1408.4442

\bibitem{Cruz07-sci} M.~Cruz, N.~Turok, P.~Vielva, E.~Martinez-Gonzalez, M.~Hobson, 2007, Science, 318, 1612-1614

\bibitem{Cruz08} M.~Cruz et al., 2008, MNRAS, 390, 913

\bibitem{Feeney12} S.~M.~Feeney, M.~C.~Johnson, D.~J.~Mortlock, H.~V.~Peiris, 2012, Phys.~Rev.~Lett.~108,~241301

\bibitem{Aguirre07} A.~Aguirre, M.~C.~Johnson, \& A.~Shomer, 2007, Phys.~Rev.~D~76,~063509.

\bibitem{Gurzadyan13} V.~G.~Gurzadyan, \& R.~Penrose, 2013, EPJP, 128, 22

\bibitem{Feeney13} S.~M.~Feeney et al., 2013,~Phys.~Rev.~D.~88~043012.

\bibitem{Li:2009sp}
  M.~Li and Y.~Wang,
  JCAP {\bf 0907}, 033 (2009)
  [arXiv:0903.2123 [hep-th]].

\bibitem{Afshordi:2010wn}
  N.~Afshordi, A.~Slosar and Y.~Wang,
  JCAP {\bf 1101}, 019 (2011)
  [arXiv:1006.5021 [astro-ph.CO]].

\bibitem{Sanchez14} J.~C.~Bueno Sanchez, 2014, Phys.~Lett.~B., 739, 269


\bibitem{Dvali:1998pa}
  G.~R.~Dvali and S.~H.~H.~Tye,
  Phys.\ Lett.\ B {\bf 450}, 72 (1999)
  [hep-ph/9812483].

\bibitem{Ma:2013xma}
  Y.~Z.~Ma, Q.~G.~Huang and X.~Zhang,
  Phys.\ Rev.\ D {\bf 87}, no. 10, 103516 (2013)
  [arXiv:1303.6244 [astro-ph.CO]].

\bibitem{Chen:2009we}
  X.~Chen and Y.~Wang,
  Phys.\ Rev.\ D {\bf 81}, 063511 (2010)
  [arXiv:0909.0496 [astro-ph.CO]].
\bibitem{Chen:2009zp}
  X.~Chen and Y.~Wang,
  JCAP {\bf 1004}, 027 (2010)
  [arXiv:0911.3380 [hep-th]].
\bibitem{Baumann:2011nk}
  D.~Baumann and D.~Green,
  Phys.\ Rev.\ D {\bf 85}, 103520 (2012)
  [arXiv:1109.0292 [hep-th]].
\bibitem{Chen:2012ge}
  X.~Chen and Y.~Wang,
  JCAP {\bf 1209}, 021 (2012)
  [arXiv:1205.0160 [hep-th]].

\bibitem{Starobinsky:1986fxa}
  A.~A.~Starobinsky,
  JETP Lett.\  {\bf 42}, 152 (1985)
  [Pisma Zh.\ Eksp.\ Teor.\ Fiz.\  {\bf 42}, 124 (1985)].

\bibitem{Sasaki:1995aw}
  M.~Sasaki and E.~D.~Stewart,
  Prog.\ Theor.\ Phys.\  {\bf 95}, 71 (1996)
  [astro-ph/9507001].

\bibitem{Lyth:2004gb}
  D.~H.~Lyth, K.~A.~Malik and M.~Sasaki,
  JCAP {\bf 0505}, 004 (2005)
  [astro-ph/0411220].

\bibitem{Chen:2007gd}
  B.~Chen, Y.~Wang and W.~Xue,
  JCAP {\bf 0805}, 014 (2008)
  [arXiv:0712.2345 [hep-th]].

\bibitem{Gurzadyan:2011ac}
  V.~G.~Gurzadyan and R.~Penrose,
  arXiv:1104.5675 [astro-ph.CO].

\bibitem{Eriksen:2011fi}
  H.~K.~Eriksen and I.~K.~Wehus,
  arXiv:1105.1081 [astro-ph.CO].

\bibitem{Kofman:2004yc}
  L.~Kofman, A.~D.~Linde, X.~Liu, A.~Maloney, L.~McAllister and E.~Silverstein,
  JHEP {\bf 0405}, 030 (2004)
  [hep-th/0403001].

\bibitem{Battefeld:2011yj}
  D.~Battefeld, T.~Battefeld, C.~Byrnes and D.~Langlois,
  JCAP {\bf 1108}, 025 (2011)
  [arXiv:1106.1891 [astro-ph.CO]].

\bibitem{Duplessis:2012nb}
  F.~Duplessis, Y.~Wang and R.~Brandenberger,
  JCAP {\bf 1204}, 012 (2012)
  [arXiv:1201.0029 [hep-th]].

\bibitem{Bousso:2000xa}
  R.~Bousso and J.~Polchinski,
  JHEP {\bf 0006}, 006 (2000)
  [hep-th/0004134].

\bibitem{Susskind:2003kw}
  L.~Susskind,
  In *Carr, Bernard (ed.): Universe or multiverse?* 2007, pp.247-266
  [hep-th/0302219].
\end{thebibliography}
\end{document}